\documentclass[reprint,nofootinbib,amsmath,amssymb,aps,floatfix]{revtex4-2}

\usepackage{bm}
\usepackage{dcolumn}

\usepackage{graphicx}
\usepackage{subcaption}
\usepackage{placeins}
\usepackage{float}

\usepackage{booktabs}
\usepackage{multirow}

\usepackage{xspace}
\usepackage{xcolor}
\usepackage[colorlinks=true,linkcolor=blue,citecolor=blue,urlcolor=blue]{hyperref}
\usepackage[font=small,labelfont=bf,textfont=it]{caption}
\usepackage{pifont}
\newcommand{\cmark}{\ding{51}}
\newcommand{\xmark}{\ding{55}}

\begin{document}

\preprint{APS/123-QED}

\title{Time-Domain Synthesis of Gravitational-Wave Detector Glitches using Class-Conditional Derivative Generative Adversarial Networks}

\author{Tom Dooney$^{1,2}$}
\author{Mees de Boer$^{2}$}
\author{Harsh Narola$^{1,2}$}
\author{Melissa Lopez$^{1,2}$}
\author{Stefano Bromuri$^{3}$}
\author{Daniel Stanley Tan$^{3}$}
\author{Chris Van Den Broeck$^{1,2}$}

\affiliation{$^1$Nikhef, Science Park 105, 1098 XG Amsterdam, The Netherlands}
\affiliation{$^2$Institute for Gravitational and Subatomic Physics (GRASP), Utrecht University, Princetonplein 1, 3584 CC Utrecht, The Netherlands}
\affiliation{$^3$Faculty of Science, Open Universiteit, Valkenburgerweg 177, 6419 AT Heerlen, The Netherlands}

\date{\today}

\begin{abstract}
Gravitational-wave detectors such as LIGO, Virgo, and KAGRA are highly sensitive instruments susceptible to numerous noise sources. 
Short-duration transient noise events, known as glitches, pose a particular challenge for data analysis pipelines as they can mimic or obscure astrophysical signals. 
We present GlitchGAN, a class-conditional generative model, built upon the Conditional Derivative GAN (cDVGAN) architecture, that is capable of synthesizing glitches directly in the time domain. 
The model is trained on high-quality reconstructions of seven common glitch types observed during LIGO's third observing run (O3): Blip, Fast Scattering, Koi Fish, Low-Frequency Burst, Scattered Light, Tomte, and Whistle.
We show that GlitchGAN generalizes effectively, learning to reproduce a diverse glitch space that is consistent with these reconstructions. 
Moreover, because the model is conditioned on glitch class, it can generate \textit{hybrid} or transitional glitch morphologies by interpolating across the class-conditioning vector after training.
GlitchGAN generates 1000 glitches in under 22 seconds on a CPU, making it suitable for large-scale glitch synthesis for detector simulations, mock data challenges, and pipeline validation.
Synthetic glitches are validated against real glitch reconstructions using the Gravity Spy classifier, widely used in the GW community for glitch classification, and an unsupervised analysis using UMAP embeddings. 
Gravity Spy classifies the majority of GlitchGAN's synthetic glitches as the correct class while the UMAP analysis shows substantial overlap between real and synthetic samples in the reduced latent space. 
To investigate residual distributional differences between real glitches unseen during training and generated synthetic glitches, we train a separate holdout GlitchGAN model on nearly the same dataset but with a small held-out split, and use a downstream convolutional neural network to distinguish these held-out real glitches from synthetic glitches produced by the holdout model.
While this classifier reliably detects synthetic samples in a clean, noise-free representation, this detectability drops substantially once real and synthetic glitches are injected into realistic detector noise, the condition under which synthetic glitches would typically be deployed for downstream applications.
Despite remaining statistically distinguishable, we show that GlitchGAN-generated glitches are practically useful: augmenting real training datasets with synthetic samples matches simple duplication of the same real data when data is abundant, and increasingly outperforms it as real data becomes scarcer.
Finally, we highlight a critical limitation of magnitude-only spectrograms: classifiers operating on magnitude $Q$-transforms can confidently misclassify physically unrealistic glitches from less robust models, underscoring the need for complementary validation methods that preserve phase information.
\end{abstract}

\maketitle


\section{Introduction}

The detection of gravitational waves (GWs) by the LIGO \cite{LIGO_paper} and Virgo \cite{VIRGOpaper} detectors has opened a new observational window on the universe, enabling measurements of compact binary coalescences and other cataclysmic astrophysical events~\cite{GWTC_3_catalog}.
The extreme sensitivity required for these detections, however, renders the instruments susceptible to instrumental and environmental noise sources~\cite{LIGO_O3_detector_characterization}.
Among these, short-duration transient noise artefacts known as \textit{glitches}~\cite{LIGO_second_third_detchar, Virgo_detchar_O3, Kagra_detchara} represent a persistent challenge for data analysis pipelines and extensive work is required to mitigate their impact, ensuring the quality of gravitational-wave data~\cite{davis2026rapiddataqualityinvestigations}.
Glitches are unmodelled noise events that exhibit diverse morphologies---ranging from impulsive millisecond-duration events to extended, dispersive features spanning several seconds---and can mimic or obscure astrophysical signals~\cite{abbott2018effects, blackburn2008lsc, abbott2016characterization}.
When a glitch overlaps with a GW signal it can significantly bias parameter estimation~\cite{pankow2018mitigation, PE_glitch_1, PE_glitch_2}, and approximately 25 of the $\sim\!90$ confident detections up to the end of O3 required some form of glitch mitigation~\cite{GWTC_3_catalog, pankow2018mitigation, BNS_1, BNS_2, BNS_BBH_2}.
The contamination of astrophysical signals by glitches will be exacerbated in next-generation detectors. Einstein Telescope \cite{ET_paper} and Cosmic Explorer \cite{Cosmic_expl} are expected to detect on the order of $10^4-10^5$ binary black hole and neutron star mergers per year~\cite{Iacovelli_2022, Kalogera_BNS}, thus giving rise to more frequent glitch-signal overlaps.

The ability to synthesize realistic glitch populations is therefore of growing importance.
Such capability enhances the realism of detector simulations and supports software injection campaigns~\cite{Abadie_2010_injection_1, Abadie_2010_injection_2, Virgo_detchar_O3}, enabling more rigorous validation of GW detection pipelines and parameter estimation algorithms.
Realistic synthetic glitches are also essential for the construction of mock data challenges (MDCs) and for data augmentation in machine learning-based glitch classification systems.
A possible approach is to model glitches mathematically or using machine learning, given \textit{a priori} knowledge of their production mechanism \cite{Tolley:2023umc, Chatterjee_2025}. 
Other works have developed models that mimic their principal characteristics using a few parameters \cite{Bondarescu:2023jcx}. 
However, such approaches either require prior knowledge of the glitch's origin, which isn't always available, or are limited to a restricted parametric form, making them difficult to scale across the full diversity of glitch morphologies.

Recently, generative models have been applied broadly within GW data analysis to address multiple challenges~\cite{Liao:2021vec, PhysRevD.110.104055}.
For glitch synthesis, Generative Adversarial Networks (GANs)~\cite{Goodfellow:2014upx} have been used to augment spectrogram-based datasets for downstream classification tasks~\cite{GAN_spec_best, Jade_Powell_paper}.
The \textit{gengli} glitch generator~\cite{GENGLI, GENGLI_2} demonstrated that a GAN trained on Blip glitches extracted from detector data using BayesWave~\cite{BayesWave}, the state-of-the-art method for reconstructing glitches from detector data, can successfully learn their underlying time-domain distribution, providing a useful framework for simulations and MDCs.
McGANn~\cite{McGinn_2021} implemented a conditional GAN capable of synthesizing five distinct waveform classes analogous to GW bursts in one model, with class interpolation enabling the generation of hybrid morphologies that blend features from the learned classes.
Powell et al.~\cite{Jade_Powell_paper} implemented generative modelling on a broader glitch class space by training separate GANs on magnitude $Q$-scan representations of each glitch class.
However, extending this approach to a single model operating directly in the time domain remains unexplored, despite its advantages in preserving phase information and enabling direct use of time-series data in simultions and MDCs.

The principal obstacle to realistic time-domain glitch synthesis has been the availability of high-quality training data.
BayesWave~\cite{BayesWave} is computationally expensive, and converges unreliably for certain glitch morphologies, making it impractical to construct the large-scale, diverse datasets that cover the broad glitch space.
The recent development of DeepExtractor~\cite{DeepExtractor_paper}---a deep learning framework for time-domain glitch reconstruction---offers fast, high-fidelity reconstructions that preserve both amplitude and phase information without imposing waveform smoothness assumptions.
DeepExtractor has been shown to yield better time-domain mismatch performance than BayesWave in a simulated glitch recovery experiment for both in-sample training classes and out-of-sample classes unseen during training (see FIG. 7 in~\cite{DeepExtractor_paper}), achieving a $10{,}000\times$ speed-up over BayesWave in realistic 1:1 comparisons.

Furthermore, DeepExtractor has been shown to be able to reconstruct real signals in both LIGO's Hanford and Livingston detectors during O3 with high fidelity without being trained on gravitational waveforms. 
DeepExtractor achieved low time-domain mismatch against the maximum-likelihood templates from standard parameter estimation workflows (see FIG. 11 in~\cite{DeepExtractor_paper}), where all signals in the experiment with single-detector matched-filter signal-to-noise ratio (SNR) $\rho_{MF}>13.5$ are reconstructed with under 10\% mismatch against maximum-likelihood templates.
Such findings support DeepExractor as a reliable tool for glitch reconstruction that is agnostic of the underlying morphology, making it suitable for generating large-scale training datasets for generative models.

\begin{figure*}[t]
\centering
\includegraphics[width=\linewidth]{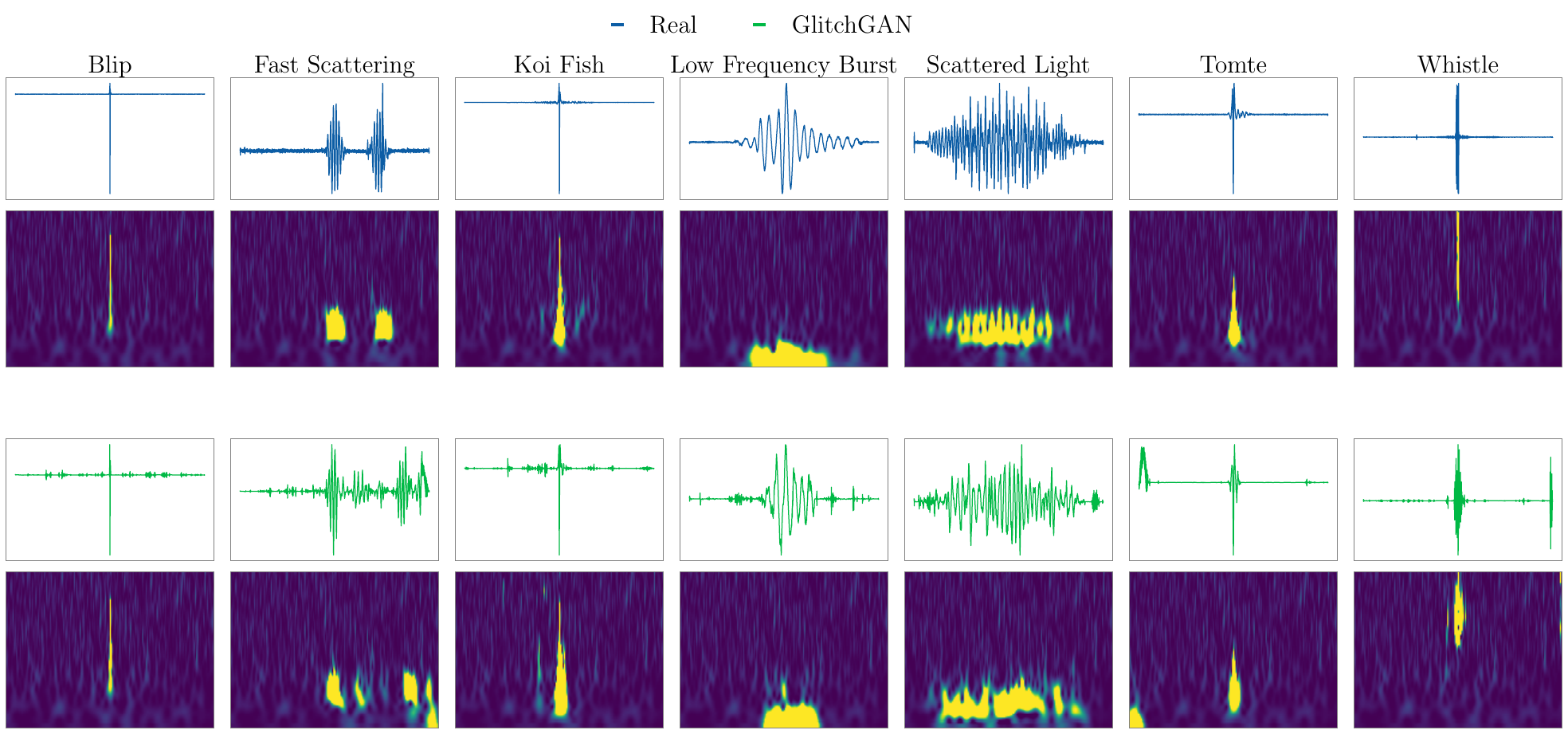}
\caption[Real and generated glitch waveforms for seven classes]{
Comparison of real and GlitchGAN-generated glitches across the seven 
glitch classes used for training, shown in both representations.
\textbf{Row 1 (Real, time domain):} DeepExtractor reconstructions of 
real glitches.
\textbf{Row 2 (Real, $Q$-transform):} Corresponding $Q$-scans of the 
same real glitches.
\textbf{Row 3 (Synthetic, time domain):} Glitches generated by 
GlitchGAN.
\textbf{Row 4 (Synthetic, $Q$-transform):} Corresponding $Q$-scans of 
the generated glitches.
Each column shows one representative sample per class.}
\label{fig:real_vs_generated_waveforms}
\end{figure*}

In this paper, we introduce GlitchGAN, a class-conditional model trained on a dataset of DeepExtractor reconstructions of seven common glitch classes observed during LIGO's O3 run: Blip, Fast Scattering, Koi Fish, Low-Frequency Burst, Scattered Light, Tomte, and Whistle.
Examples of the glitches used for training and the corresponding GlitchGAN outputs are shown in Figure~\ref{fig:real_vs_generated_waveforms}.
GlitchGAN inherits the Conditional Derivative GAN (cDVGAN) architecture~\cite{cdvgan, dooney2022dvgan}, which features an auxiliary discriminator applied to the time derivative of signals, providing additional regularization that improves training stability and sample fidelity compared to standard GAN counterparts.

We validate the quality of the DeepExtractor training data using the Gravity Spy classifier~\cite{Zevin_2017, Glanzer_2023, GspyO4} and unsupervised t-SNE \cite{TSNE} and UMAP embeddings~\cite{McInnes_2018}, which have been used in multiple studies to characterize the morphological distribution of glitches in LIGO data~\cite{ferreira2025analysisligoglitchesusing, Ferreira_2025, deshmukh2026soundnoiseleveraginginductive}.
The evaluation establishes a realistic baseline against which synthetic glitches can be assessed.
Applying the same evaluation to GlitchGAN's outputs, we show that the model captures much of the morphological diversity across all seven classes, with substantial overlap between real and generated distributions in the t-SNE and UMAP latent spaces and with the majority of synthetic glitches correctly identified by Gravity Spy.

Furthermore, we train a dedicated convolutional neural network (CNN) binary classifier to distinguish real glitches held out from training from GlitchGAN-generated samples. While this classifier reliably detects synthetic samples in the clean, noise-free representation used for training (95.7\% accuracy), a simple linear logistic regression baseline classifier performs substantially worse (73.6\% accuracy), indicating that the residual difference is largely nonlinear rather than a significant morphological mismatch.
Injecting both populations into realistic detector noise reduces this detectability further (66.0\% CNN accuracy), and we show that despite remaining statistically distinguishable, GlitchGAN's synthetic glitches are practically useful: augmenting real training datasets with synthetic samples matches simple duplication of the same real data when data is abundant, and increasingly outperforms it as real data becomes scarcer -- remaining safe where duplication turns harmful, and substantially improving downstream multi-class classification accuracy in the most severely data-scarce regime.

We also demonstrate that GlitchGAN can interpolate between classes, producing novel hybrid morphologies that mix features from the learned classes.
Finally, we highlight a fundamental limitation of magnitude-only time-frequency representations as validation tools: when applied to physically unrealistic glitches from a diffusion-based model~\cite{diffusionOriginal}, Gravity Spy assigns correct class labels with high confidence despite clear inconsistency in the time domain, because the $Q$-transform discards phase information entirely.
This underscores the necessity of complementary, phase-preserving validation methods.

The remainder of the paper is organized as follows:
Section~\ref{sec:data} describes the preparation of our training dataset.
Section~\ref{sec:methods} details the GlitchGAN architecture, training procedure and evaluation methodology.
Section~\ref{sec:results} presents results across multiple metrics.
Finally, Section~\ref{sec:discussion} discusses our findings and opportunities for future work.

\vspace{-1mm}
\section{Data}
\label{sec:data}

\vspace{-1mm}

\subsection{Data curation}

We construct our training dataset using DeepExtractor reconstructions of real LIGO glitches.
To ensure high-quality, morphologically representative samples, we apply strict selection criteria:

\begin{enumerate}
\item
Gravity Spy classification confidence $\sigma_{\text{GS}} \geq 0.9$, ensuring that the reconstructed glitches are representative of well-characterized classes.

\item
Signal-to-noise ratio (SNR) $\geq 15$ (sufficient amplitude to minimize reconstruction artifacts)

\item
Detectors: LIGO Hanford (H1) and Livingston (L1)

\item
Observing runs: O3a and O3b


\end{enumerate}

We include seven glitch classes in the training of GlitchGAN, namely; Blip, Fast Scattering, Koi Fish, Low-Frequency Burst, Scattered Light, Tomte, and Whistle.

\textit{Blip} glitches have a characteristic morphology of a symmetric teardrop shape in time-frequency in the range $[30, 250]\,$Hz with short-durations, $\sim 0.04\,$s \cite{abbott2016characterization}. 
They appear in both Hanford and Livingston, as well as Virgo and GEO $600$ \cite{Cabero:2019orq}. 
Due to their abundance and form, these glitches hinder both the unmodeled burst and modelled CBC searches, with particular emphasis on compact binaries with large total mass, highly asymmetric component masses, and spins anti-aligned with the orbital angular momentum~\cite{abbott2018effects}.

\textit{Fast Scattering} glitches appear as short-duration arches ($\sim 0.2 - 0.3\,$s) in the frequency range $[20, 60] \,$Hz. These glitches are strongly correlated with ground motion in range $[0.1 -0.3] \,$Hz and $[1-6] \,$Hz, which in turn is associated with thunderstorms and human activity near the detector. Note that this class was added during O3, and it is more abundant in Livingston than Hanford, due to differences in ground motion and detector sensitivity \cite{Soni:2021cjy}.

\textit{Koi Fish} glitches get their name from their frontal resemblance to Koi Fishes. They are also similar to \textit{Blips} but typically feature a higher SNR, spanning the frequency range of $[20, 1000] \,$Hz.

\textit{Low-Frequency Bursts} are short-duration  glitches ($\sim 0.25\,$s) in the frequency range $[10 - 20]\,$ Hz with a distinctive blob shape. These occurrences were prevalent in Livingston data during O1 and Hanford data in O3a.

\textit{Scattered Light} glitches have longer duration harmonics ($\sim2.0 - 4.0\,$s), and in the time-frequency domain, they appear as arches being often stacked on top of each other. These glitches are problematic since their frequency content lies in the band of interest of GW astrophysical events. In O3, they were found to be coupled with the relative motion between the optical suspension system’s end test-mass chain and the reaction-mass chain \cite{Soni:2021cjy}.

\textit{Tomte} glitches are also short-duration ($\sim 0.25\,$s) with characteristic triangular morphology.

\textit{Whistles} have a characteristic V, U or W shape at higher frequencies with typical durations $\sim 0.25\,$s. They are caused when radio-frequency signals beat with the voltage-controlled oscillators \cite{Nuttall:2015dqa}.

These seven classes collectively span wide time and frequency ranges, encompassing much of the morphological variability relevant to transient instrumental noise.
While it is a common glitch class, we exclude the Extremely Loud class because its characterization is dominated by exceptionally high SNR rather than distinctive morphology, making it a poor target for generative modeling.


\subsection{Dataset preprocessing}

We reconstruct with DeepExtractor a 2-second window centered on the glitch trigger time at a sampling rate of 4,096~Hz, yielding 8,192 samples per glitch waveform.
Our goal was to obtain approximately 5,000 samples per class (1,250 per detector per observing run) under the selection criteria above.
Unfortunately, data availability constraints limited acquisition for some classes.
Fast Scattering yielded only 2,774 samples, and Low-Frequency Burst provided 4,316 samples.
The total curated dataset contained 31,511 samples.

To mitigate training instabilities from class imbalance, we applied bootstrap resampling (sampling with replacement) to oversample all classes to 5,000 samples each, yielding a balanced training set of 35,000 total samples.
Dataset composition by detector, observing run, and class is detailed in Table~\ref{tab:glitch_pivot_summary}.
Before training, all reconstructed time series are normalized to the range $[-1, 1]$, followed by subtracting the mean.
This normalization constrains the generative model to operate at a single effective SNR scale, but generated samples can be rescaled post-generation to any desired SNR for downstream applications.

\begin{table}[ht]
    \centering
    \caption{Composition of the curated glitch dataset. Number of samples per class, detector (H1: Hanford, L1: Livingston), and observing run (O3a, O3b). Classes are bootstrapped to 5000 samples each for balanced training.}
    \label{tab:glitch_pivot_summary}
    \begin{tabular}{l rrrr}
    \toprule
    \multirow{2}{*}{Glitch Class} & \multicolumn{2}{c}{H1} & \multicolumn{2}{c}{L1} \\
    \cmidrule(lr){2-3} \cmidrule(lr){4-5}
    & O3a   & O3b   & O3a   & O3b   \\
    \midrule
    Blip                  & 1246 & 1247 & 1245 & 1243 \\
    Fast Scattering      & 143  & 143  & 1244 & 1244 \\
    Koi Fish             & 1247 & 1248 & 1243 & 1247 \\
    Low Freq. Burst     & 1246 & 1246 & 912  & 912  \\
    Scattered Light      & 1236 & 1243 & 1246 & 1248 \\
    Tomte                 & 1031 & 1031 & 1247 & 1246 \\
    Whistle               & 1232 & 1232 & 1231 & 1232 \\
    \bottomrule
    \end{tabular}
\end{table}

\section{Methods}
\label{sec:methods}

\subsection{Model architecture: cDVGAN}

Generative adversarial networks (GANs)~\cite{Goodfellow:2014upx}, wherein a generator and discriminator are trained in an adversarial min-max game, have established themselves as a powerful framework for learning complex high-dimensional data distributions.
Wasserstein GANs (WGANs)~\cite{wGAN_paper} improve training stability by replacing the Jensen--Shannon divergence with the Wasserstein-1 distance, and the gradient penalty variant (WGAN-GP)~\cite{wGAN_GP_paper} further stabilizes optimization by enforcing a Lipschitz constraint on the discriminator.
Conditional extensions~\cite{Mirza_2014} allow the generator to be steered towards a target class, enabling multi-class synthesis within a single model.

cDVGAN~\cite{cdvgan} is a conditional WGAN-GP that augments the standard two-player adversarial game with an auxiliary discriminator applied to the first-order time derivative of signals.
Here we summarise the key components; full layer-by-layer architecture details are shown in Appendix~\ref{sec:cDVGAN_Architecture}.

\subsubsection{Training objective}

Under the WGAN-GP framework, the loss for discriminator $D$ is
\begin{equation}\label{eq:discriminator_loss}
    L_D = \mathbb{E}_{\hat{x}}[D(\hat{x},c)] - \mathbb{E}_{x}[D(x,c)] + \lambda\,\mathbb{E}_{\tilde{x}}\!\left[\left(\|\nabla_{\tilde{x}} D(\tilde{x},c)\|_2 - 1\right)^2\right],
\end{equation}
where $x$ and $\hat{x}$ are real and generated samples respectively, $\tilde{x}$ is a random interpolation between them, and $\lambda = 10$ is the gradient-penalty weight.
The generator is updated to minimize
\begin{equation}
    L_G = -\mathbb{E}_{\hat{x}}[D(\hat{x},c)].
\end{equation}

In cDVGAN a second discriminator $D_\partial$ operates on the first-order derivative $\dot{x} = dx/dt$.
The combined generator loss is
\begin{equation}\label{eq:combined_loss}
    L_G^{\text{cDVGAN}} = -\frac{1}{2}\left(\mathbb{E}_{\hat{x}}[D(\hat{x},c)] + \mathbb{E}_{\dot{\hat{x}}}[D_\partial(\dot{\hat{x}},c)]\right),
\end{equation}
where both discriminators contribute equally.
Each discriminator is updated five times per generator update.

\subsubsection{Generator}

The generator $G$ maps a noise vector $z \sim \mathcal{N}(0, I_{100})$ and a one-hot class vector $c \in \mathbb{R}^7$ to a time series of length 8,192.
The class vector is first projected to a 32-dimensional embedding and concatenated with $z$, then passed through a dense layer and reshaped before five successive upsample-and-convolve blocks with batch normalization and ReLU activations expand the sequence to the target length.

\subsubsection{Discriminators}

Both $D$ and $D_\partial$ are 1D convolutional networks that map an input sequence and a class vector to a scalar critic value.
Class conditioning uses projection~\cite{Mirza_2014}: a learned class embedding is dot-producted with the discriminator's feature vector and added to the scalar output.
$D$ operates directly on the 8192-sample waveform; $D_\partial$ operates on the 8191-sample first-order difference sequence $\dot{x} = x_{t+1} - x_t$.
The derivative discriminator provides additional regularization by penalizing unphysical high-frequency artefacts in generated waveforms~\cite{dooney2022dvgan}.

\subsection{Training procedure\label{sec:training_procedure}}

The model was trained for 500 epochs with a batch size of 64 using the RMSprop optimizer and a learning rate of $10^{-4}$ for all network components.
Training was performed on a NVIDIA A100 GPU (approximately 12 hours).
Although training was conducted for the full 500 epochs, the final GlitchGAN configuration uses the model weights saved at epoch 210.


The selection of epoch 210 was based on a combination of qualitative and quantitative assessments. Generated glitches were inspected throughout training, and epoch 210 was found to produce the most realistic glitch morphologies. This choice is further supported by the generator loss evolution (Appendix \ref{sec:cDVGAN_Loss}), which exhibits relatively stable behaviour around this epoch compared to other stages of training. Additionally, the UMAP analysis presented in Section \ref{sec:glitchgan_umap} shows improved overlap between the distributions of real and generated glitches at epoch 210 relative to models saved at epochs 100, 200, 300, 400, and 500.
 
No hyperparameter tuning was performed for this dataset; the architecture and hyperparameters are similar to those validated in~\cite{cdvgan}, outside of the increased dimensionality of the class vector, allowing us to assess out-of-the-box generalization to this more complex and realistic glitch population.
The implementation uses TensorFlow~\cite{tensorflow2015-whitepaper} and is publicly available at \url{https://github.com/tomdooney95/glitchgan}.

\subsection{Comparison with a diffusion-based model}

As a comparison against our cDVGAN architecture in GlitchGAN, we also trained another type of generative model on our dataset known as a latent Diffusion Transformer (DiT), which was adapted for 1D time series.
Diffusion models~\cite{Ho2020DDPM, diffusionOriginal, ho2022classifierfree} learn to generate data by explicitly modelling the process of iteratively adding and then removing noise from samples drawn from a simple prior distribution.
They have become increasingly popular in recent years for their ability to generate realistic synthetic data.
Full-resolution diffusion proved computationally prohibitive, so we first trained a convolutional autoencoder to compress inputs from $(8192,1)$ to a latent space of $(256,16)$, reducing computational cost by approximately three orders of magnitude while preserving essential morphological structure.
The DiT was then trained in this latent space, with generated samples decoded back to the time domain for evaluation.
This model serves primarily to illustrate the limitation of magnitude-only validation metrics, as discussed in Section~\ref{sec:results}.

\subsection{Evaluation methodology}



We evaluate both the training data (DeepExtractor reconstructions) and GlitchGAN-generated glitches using two complementary approaches: classification with the Gravity Spy classifier and visualization of low-dimensional embeddings via t-SNE and UMAP. 
For DeepExtractor, this establishes a quantitative baseline confirming the training data is morphologically representative; for GlitchGAN, it assesses how faithfully the model reproduces the underlying glitch distributions.
We further introduce a downstream binary classification test, in which a dedicated convolutional network attempts to distinguish held-out real glitches from GlitchGAN-generated ones, and repeat it with both real and generated glitches injected into realistic detector noise at per-sample SNRs drawn from the empirical distribution of each class. 
Finally, we evaluate a practical downstream application: whether, under the same realistic noise conditions, GlitchGAN-generated glitches can augment real training sets across a range of data scarcity to improve seven-class glitch classification accuracy.

\subsubsection{Gravity Spy classification\label{sec:gspy_classification_methodology}}

We reconstruct with DeepExtractor 100 randomly selected glitches from the Gravity Spy dataset according to the criteria in Section \ref{sec:data}, drawing 25 samples per detector (H1, L1) per observing run (O3a, O3b).
Since Gravity Spy classifies glitches surrounded by detector background noise i.e. $g(t) + n(t)$, each DeepExtractor reconstruction $\hat{g(t)}$ is injected into a segment of `quiet' Hanford detector noise from O3 between 1262540000 and 1262540040 GPS time prior to reclassification (representing $n(t)$), using the \texttt{fetch\_open\_data} method from the package gwpy~\cite{gwpy}.
This stretch of data was visually inspected using $Q$-scans to confirm the absence of glitches or other significant noise features.

We follow a similar approach to evaluate the GlitchGAN-generated synthetic glitches, injecting them into the same noise segments before classification. 
The only difference is that we have to explicitly rescale the generated glitches to a fixed SNR to ensure they are sufficiently loud for classification, since the generator is trained on normalized waveforms and does not learn an explicit SNR distribution. 
We fix the SNR of each glitch class according to the mean SNR of the respective class considering the selection criteria described in Section \ref{sec:data}.
The mean SNR values used for each class are shown in Table~\ref{tab:injection_snr}.

Since each per-class accuracy estimate is derived from only $n=100$ trials, we report 95\% confidence intervals alongside all classification accuracies (for both DeepExtractor and GlitchGAN) to quantify sampling uncertainty. 
Because per-class accuracy is a binomial proportion (the fraction of $n$ trials classified into the correct class), we compute intervals using the Wilson score method~\cite{Wilson01061927}, which remains bounded within $[0,1]$ and maintains accurate coverage even for small sample sizes or accuracies near the extremes (e.g. 100\%, as with the Low Frequency Burst class). 
For the mean Gravity Spy confidence scores, which are continuous and often skewed toward high values rather than normally distributed, we instead report 95\% confidence intervals obtained via a percentile bootstrap: resampling the observed confidence values with replacement 10{,}000 times and taking the 2.5th and 97.5th percentiles of the resulting distribution of means.

\begin{table}[h]
\centering
\caption{Mean SNR per glitch class used for injections in the GlitchGAN evaluation.}
\label{tab:injection_snr}
\begin{tabular}{l r}
\toprule
Glitch Class & Mean SNR \\
\midrule
Blip              & 29.8  \\
Fast Scattering   & 36.4  \\
Koi Fish          & 187.2 \\
Low Frequency Burst & 40.3 \\
Scattered Light   & 31.5  \\
Tomte             & 25.4  \\
Whistle           & 27.1  \\
\bottomrule
\end{tabular}
\end{table}

\subsubsection{Dimensionality reduction}
To visualize and compare the high-dimensional feature space of real and synthetic glitches, we use two complementary nonlinear dimensionality reduction methods: \textit{t-distributed Stochastic Neighbor Embedding} (t-SNE)~\cite{TSNE} and \textit{Uniform Manifold Approximation and Projection} (UMAP)~\cite{McInnes_2018}. While \textit{Principal Component Analysis} (PCA) captures global variance through linear projections, it cannot resolve the nonlinear structure present in morphologically diverse glitch populations; t-SNE and UMAP are instead designed to preserve local cluster structure in the embedding. t-SNE embeds data by minimizing the Kullback--Leibler divergence between probability distributions encoding pairwise similarities in the original and reduced spaces; it excels at revealing local cluster structure but can distort global geometry. UMAP constructs a nearest-neighbour graph in the high-dimensional space and optimises a low-dimensional counterpart to preserve both local and global topological structure, scaling efficiently to the dataset sizes used here.

For DeepExtractor's reconstructions, we aim to verify that the reconstructed glitches cluster according to their correct Gravity Spy class in an unsupervised manner, confirming that the reconstructions preserve the morphological characteristics recognized by Gravity Spy.
For GlitchGAN-generated synthetic glitches, we evaluate whether they cluster with the real reconstructions of their respective classes (using the same UMAP embedding for real and synthetic samples), indicating that the model has successfully learned to reproduce the underlying morphological distribution of each class.
The hyperparameters and seeds used for t-SNE and UMAP analyses are detailed in Appendix~\ref{sec:tsne_umap_params}.

\subsubsection{Downstream classification of real vs. synthetic glitches\label{sec:downstream_classification_methods}}

To evaluate whether GlitchGAN-generated glitches are distinguishable from real DeepExtractor reconstructions, we train a binary CNN classifier to discriminate between the two populations. 
To this end, we train a new GlitchGAN model on the same dataset, but holding out 10\% of the data for the downstream classification, stratifying the split by class. 
This amounts to 28{,}360 real training samples across the seven glitch classes, which we rebalance to 5{,}000 samples per class (35{,}000 total) following the same procedure used for the main model, and 3{,}151 held-out real samples that are never seen during training of this holdout GlitchGAN. 
This holdout model is trained separately from, and does not replace, the main GlitchGAN model reported throughout this work.

Once training of the holdout model is complete, we use its generator to produce one synthetic glitch for each of the 3{,}151 held-out real glitches, conditioned on the same class label, giving a balanced evaluation pool of 6{,}302 samples (3{,}151 real, 3{,}151 generated). 
We evaluate using the holdout generator's checkpoint at epoch 210, similar to the main GlitchGAN model rather than the final epoch of training.

We use a 1D CNN with three convolutional layers followed by two fully-connected layers to perform the classification, deliberately kept simpler than and architecturally independent of GlitchGAN's own discriminator, so that it constitutes a genuinely independent check rather than a re-run of the adversarial training objective. 
The full architecture is given in Appendix~\ref{sec:downstream_classifier_architectures}. 

The balanced real/generated pool is split 70/10/20 into training, validation, and test sets, stratified jointly by class and by real/fake domain -- giving 4{,}412 training samples, 632 validation samples, and 1{,}258 test samples, with real and generated samples represented in equal number within every class and every split (Table~\ref{tab:clf_split_breakdown}). 
Prior to classification, both real and generated samples are scaled identically via the same per-sample min-max and mean-subtraction procedure used to prepare GlitchGAN's training data, so the classifier cannot exploit a trivial normalization mismatch between the two domains. 
As an additional check on the strength of the classification signal, we also fit a logistic regression baseline model directly on the raw waveform, with its L2 regularization strength selected via the validation split, to establish how much of any discriminating information is accessible to a model with no capacity for nonlinear feature extraction. We report test accuracy together with its 95\% Wilson score confidence interval, computed using the same procedure described in Section~\ref{sec:gspy_classification_methodology}, for both classifiers.

\begin{table}[h]
\centering
\caption{Per-class breakdown of the held-out real and GlitchGAN-generated glitches used in the downstream real-vs-generated classification test. The 6{,}302-sample pool (3{,}151 real, 3{,}151 generated) is split 70/10/20 into training, validation, and test sets, stratified jointly by class and by real/fake domain, so real and generated samples are represented in equal number within every class and every split.}
\label{tab:clf_split_breakdown}
\begin{tabular}{lccccc}
\toprule
Class & Real & Generated & Train & Val & Test \\
\midrule
Blip                 & 498  & 498  & 698  & 100 & 198  \\
Fast Scattering      & 277  & 277  & 388  & 56  & 110  \\
Koi Fish             & 498  & 498  & 698  & 100 & 198  \\
Low Frequency Burst  & 432  & 432  & 604  & 86  & 174  \\
Scattered Light      & 497  & 497  & 696  & 100 & 198  \\
Tomte                & 456  & 456  & 638  & 92  & 182  \\
Whistle              & 493  & 493  & 690  & 98  & 198  \\
\midrule
Total                & 3151 & 3151 & 4412 & 632 & 1258 \\
\bottomrule
\end{tabular}
\end{table}

Having established this experiment in a clean, noise-free representation, we additionally repeat the experiment with real and generated glitches injected into simulated Gaussian detector noise, more closely resembling how these samples would be encountered in practice.\label{sec:noise_injection_methodology}

We draw a coloured noise realization from the LIGO design power spectral density (PSD) $S_n(f)$ and whiten it. 
A new, independent noise realization is drawn for every injected sample, rather than reusing a single shared segment, so that no particular realization's random fluctuations can act as a feature shared across the whole dataset.

Each glitch, real or generated, is independently rescaled to a target SNR $\rho$ before injection according to
\begin{equation}\label{eq:snr_flat}
    \rho = \lVert h \rVert_2 = \left(\sum_t h(t)^2\right)^{1/2} ,
\end{equation}
where $h(t)$ is the whitened-frame glitch waveform; because our whitening procedure gives noise with unit variance per time sample, the optimal matched-filter SNR reduces exactly to this time-domain $\ell_2$ norm.


Rather than injecting every sample of a given class at that class's single, fixed mean SNR (as in Section~\ref{sec:gspy_classification_methodology}), each individual sample here is assigned its own target SNR, drawn independently from the empirical distribution of real per-class SNR values. 
We construct, for each class, an empirical pool of observed SNR values from the Gravity Spy glitch catalog, applying the same selection criteria used to construct the training dataset (Section~\ref{sec:data}). 
The resulting empirical SNR distributions for all seven classes are shown in Appendix~\ref{sec:snr_distributions}.
For each injected sample, a target SNR is drawn with replacement from its class's empirical pool; both real and synthetic domains draw from the identical per-class distribution, so SNR itself cannot reveal a sample's origin to a downstream classifier. 

\subsubsection{Multi-class classification and data augmentation\label{sec:multiclass_methods}}

As a practical downstream application, we evaluate whether GlitchGAN-generated glitches are useful for augmenting a real training set in a seven-class glitch classification task, particularly when real data is scarce, injecting real and generated glitches into realistic detector noise following the same procedure as Section~\ref{sec:downstream_classification_methods}.

We use the same held-out real glitch set as in Section~\ref{sec:downstream_classification_methods} (3{,}151 samples across seven classes, never seen during training of the holdout GlitchGAN), and a seven-class softmax variant of the binary classifier described above: identical convolutional backbone (see Section~\ref{sec:downstream_classifier_architectures}), but with a final seven-unit softmax layer and a sparse categorical cross-entropy loss in place of the single sigmoid output.


To evaluate the effect of real-data scarcity on the value of synthetic augmentation, we construct training pools at three scarcity levels, all stratified proportionally by class from the held-out real set, using train/validation/test splits of 70/10/20, 10/20/70, and 3/20/77 respectively: an abundant pool (the full held-out set at its natural split, giving 2{,}206 real training samples total, $\sim$194--349 per class), a moderately scarce pool ($\sim$10\% train fraction, giving 316 real training samples total, ranging from 28 to 50 per class), and a severely scarce pool ($\sim$3\% train fraction, giving 95 real training samples total, ranging from 8 for Fast Scattering to 15 for the larger classes). In each case, the validation and test sets are drawn from the same held-out real set at the stated split fractions and used identically across every training configuration described below for direct comparability.

For each scarcity level, we compare four training-set compositions at two augmentation-target scales; each pool's own majority-class count, and an aggressive target of 5{,}000 samples per class, all injected into realistic detector noise and evaluated against the same held-out, noise-injected real validation and test sets:
\begin{enumerate}
\item \textbf{Unaugmented}: the real unbalanced training pool as-is.
\item \textbf{Bootstrap-duplicated}: the same real training pool, balanced via oversampling with replacement to the target scale (majority class count or 5{,}000); a control that isolates whether simply having a larger, but entirely non-novel, training set explains any observed improvement.
\item \textbf{Augmented}: the real training pool, balanced via supplementation with GlitchGAN-generated samples (conditioned on the same class label) to reach the same target scale as above.
\item \textbf{Synthetic-only}: GlitchGAN-generated samples only, at each target scale, with no real training data at all.
\end{enumerate}

\section{Results}
\label{sec:results}

\begin{figure*}[t]
\centering
\includegraphics[width=0.9\linewidth]{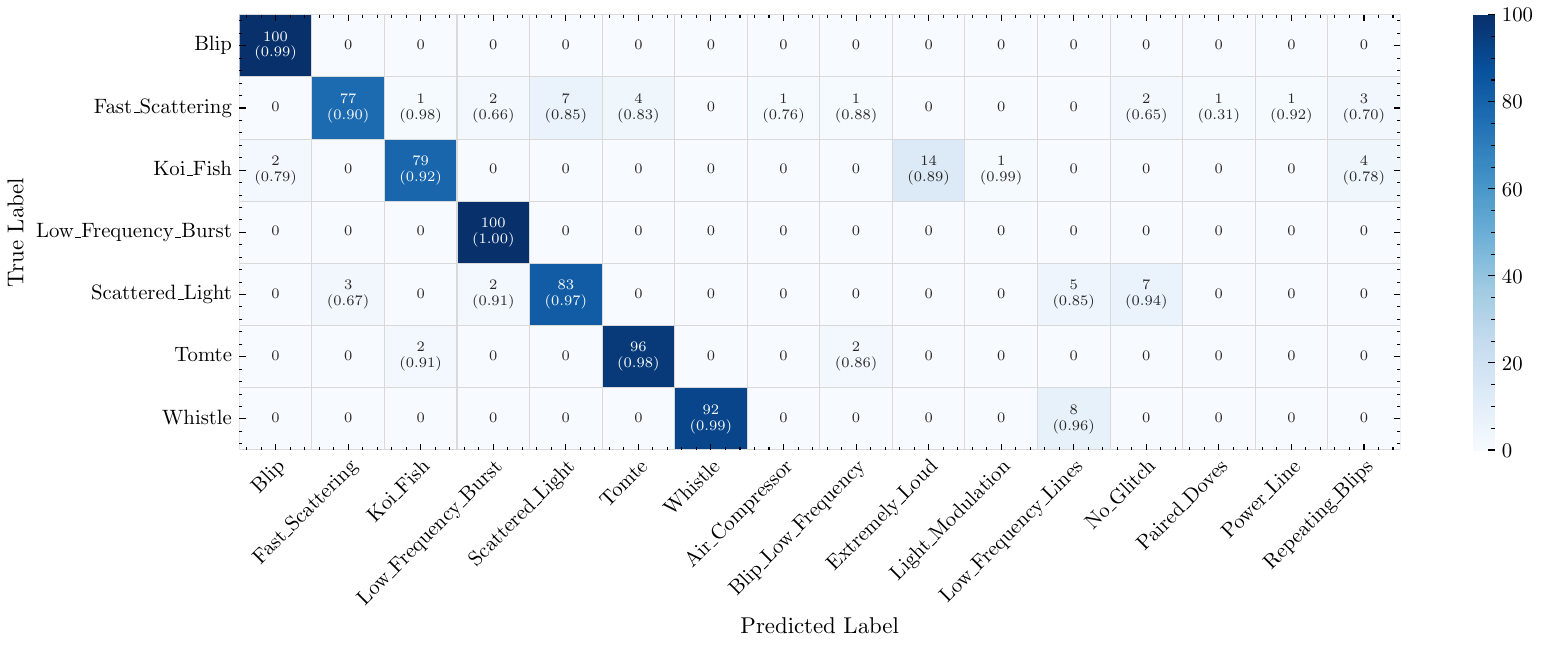}
\caption{Confusion matrix for Gravity Spy classification of 100 DeepExtractor reconstructions per class (700 total). Each cell shows the number of predictions with the mean Gravity Spy confidence in brackets.}
\label{fig:de_confusion}
\end{figure*}

\subsection{Validation of DeepExtractor training data}
\label{sec:deepextractor_validation}

\subsubsection{Classification evaluation using Gravity Spy}

Table~\ref{tab:de_ci} summarizes the results of classifying 700 DeepExtractor reconstructions (100 per class) with Gravity Spy.
Of the 700 reconstructed glitches, 627 are correctly identified by Gravity Spy, corresponding to $89.6\%$ overall accuracy (95\% Wilson CI [87.1, 91.6]\%).
Correctly classified reconstructions achieve a mean confidence of $96.8\%$ (95\% bootstrap CI [96.0, 97.5]\%), confirming that DeepExtractor effectively preserves the morphological characteristics recognized by Gravity Spy.
Misclassified samples show lower confidence of $84.8\%$, suggesting that classifier confidence may serve as a useful quality veto.

\begin{table}[h]
\centering
\caption{Gravity Spy classification results for DeepExtractor reconstructions ($n=100$ samples per class, 700 total). Accuracy intervals use the Wilson score method; confidence intervals use a percentile bootstrap (10{,}000 resamples).}
\label{tab:de_ci}
\resizebox{\linewidth}{!}{
\begin{tabular}{lcccc}
\toprule
\textbf{Class} & \textbf{Accuracy} & \textbf{95\% Wilson CI} & \textbf{Mean conf.} & \textbf{95\% bootstrap CI} \\
\midrule
Blip                 & 100.0\% & [96.3, 100.0] & 0.993 & [0.988, 0.997] \\
Fast Scattering      &  77.0\% & [67.8, 84.2]  & 0.898 & [0.863, 0.930] \\
Koi Fish             &  79.0\% & [70.0, 85.8]  & 0.916 & [0.885, 0.944] \\
Low Frequency Burst  & 100.0\% & [96.3, 100.0] & 0.998 & [0.997, 1.000] \\
Scattered Light      &  83.0\% & [74.5, 89.1]  & 0.970 & [0.954, 0.982] \\
Tomte                &  96.0\% & [90.2, 98.4]  & 0.982 & [0.963, 0.996] \\
Whistle              &  92.0\% & [85.0, 95.9]  & 0.993 & [0.982, 0.999] \\
\midrule
Overall (700)        &  89.6\% & [87.1, 91.6]  & 0.968 & [0.960, 0.975] \\
\bottomrule
\end{tabular}
}
\end{table}

Table~\ref{tab:de_ci} shows that class-level accuracy varies considerably.
All Blip and Low-Frequency Burst glitches are correctly classified, while Fast Scattering and Koi Fish achieve the lowest accuracies at 77\% and 79\%, respectively.
Figure~\ref{fig:de_confusion} shows the full confusion matrix.
Since Gravity Spy supports additional classes beyond our seven training classes, the matrix is asymmetric.
Most misclassifications occur between morphologically similar classes: among the 21 misclassified Koi Fish glitches, 14 were labelled as Extremely Loud, consistent with their exceptionally high SNRs (220--1270) overwhelming finer morphological details.
The 23 misclassified Fast Scattering samples are distributed across several related classes (Scattered Light, Tomte, Repeating Blips), reflecting the known morphological similarity between low-frequency scattering noise types.
Overall, the misclassification patterns are morphologically explainable and do not indicate systematic failure of the reconstruction.

\subsubsection{Unsupervised clustering with t-SNE and UMAP}

\begin{figure*}[tp]
\centering
\begin{subfigure}{0.45\textwidth}
    \centering
    \includegraphics[width=\linewidth,height=0.26\textheight,keepaspectratio]{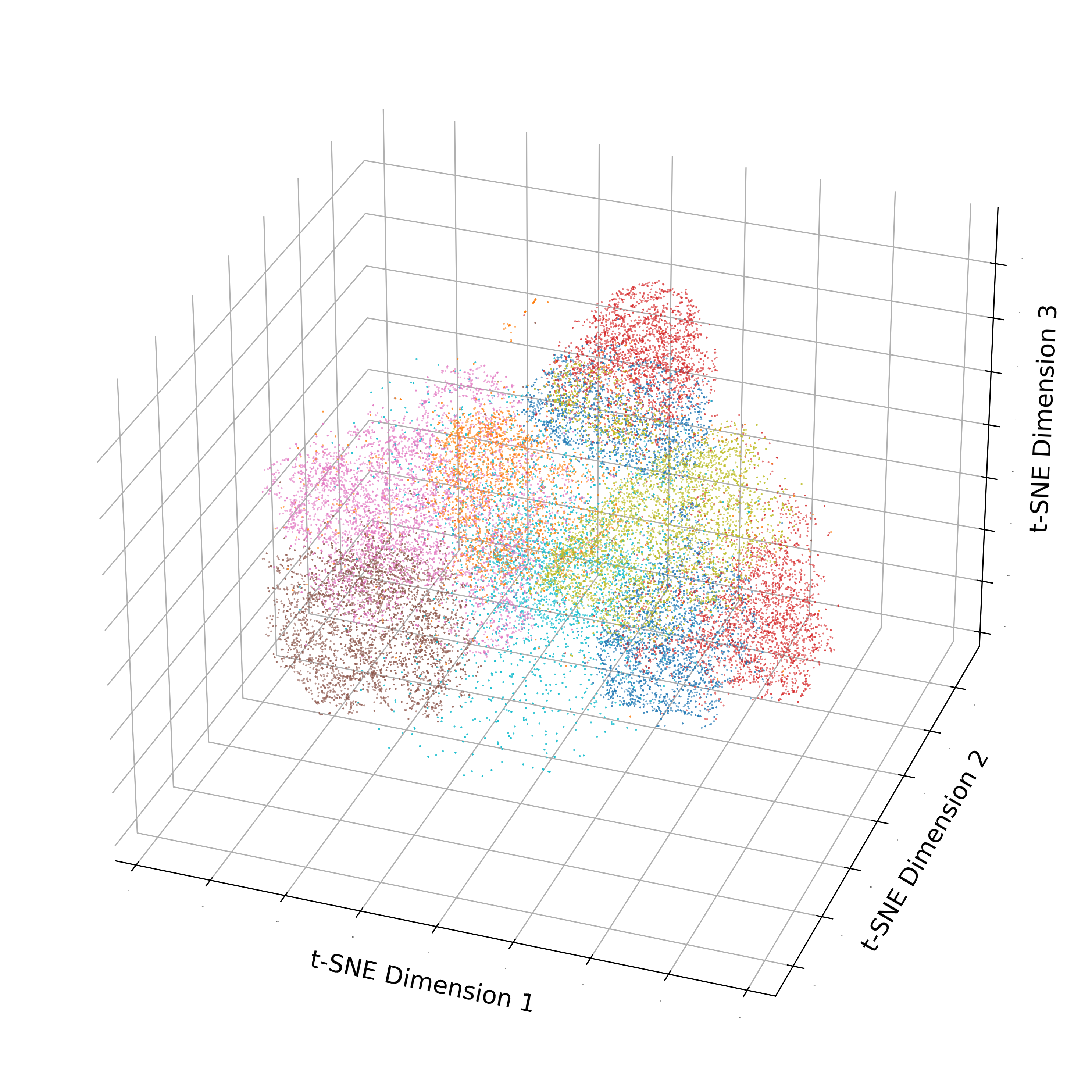}
\end{subfigure}\hfill
\begin{subfigure}{0.45\textwidth}
    \centering
    \includegraphics[width=\linewidth,height=0.26\textheight,keepaspectratio]{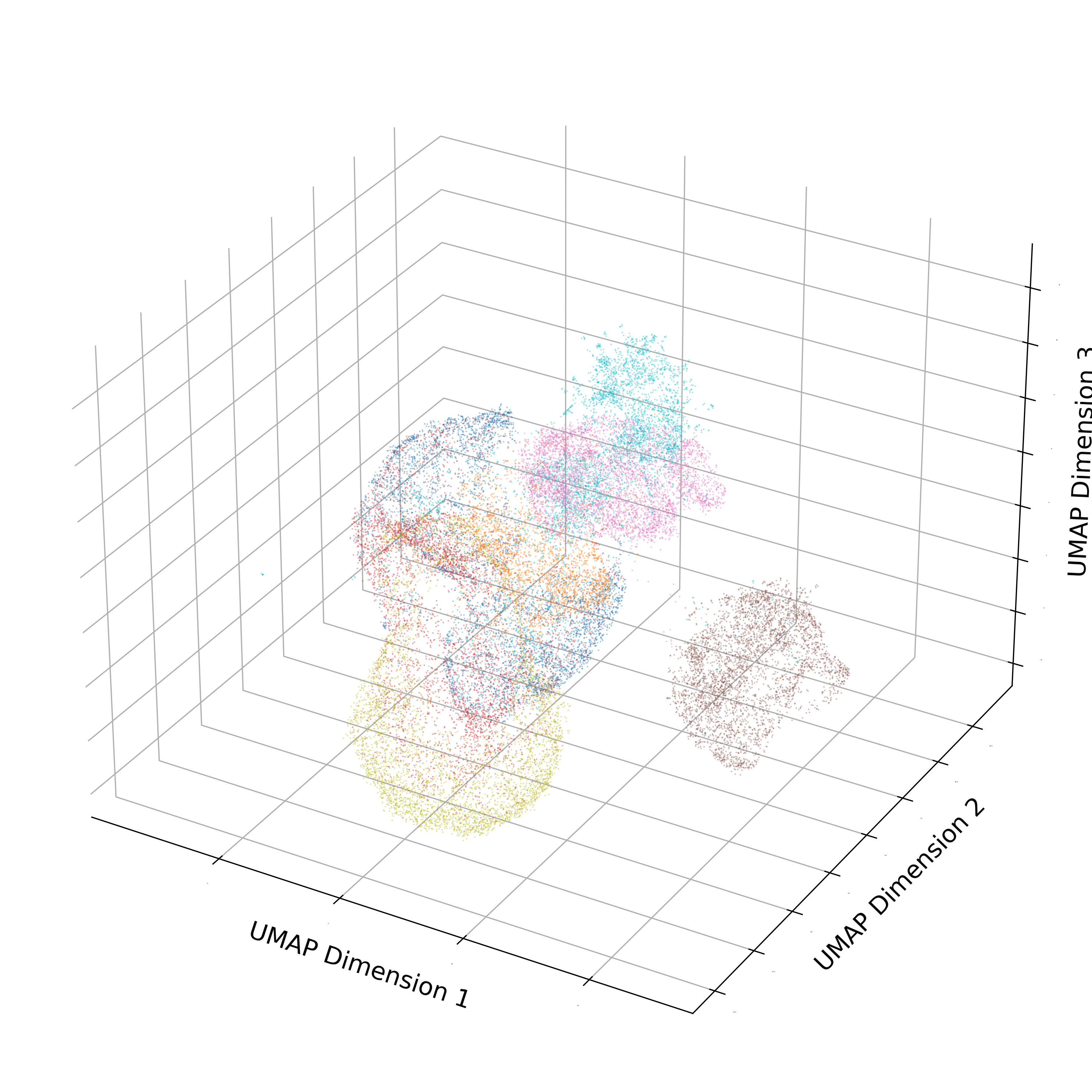}
\end{subfigure}
\vspace{-1.5em}
\begin{subfigure}{0.45\textwidth}
    \hspace*{-1cm}
    \includegraphics[width=\linewidth,height=0.26\textheight,keepaspectratio]{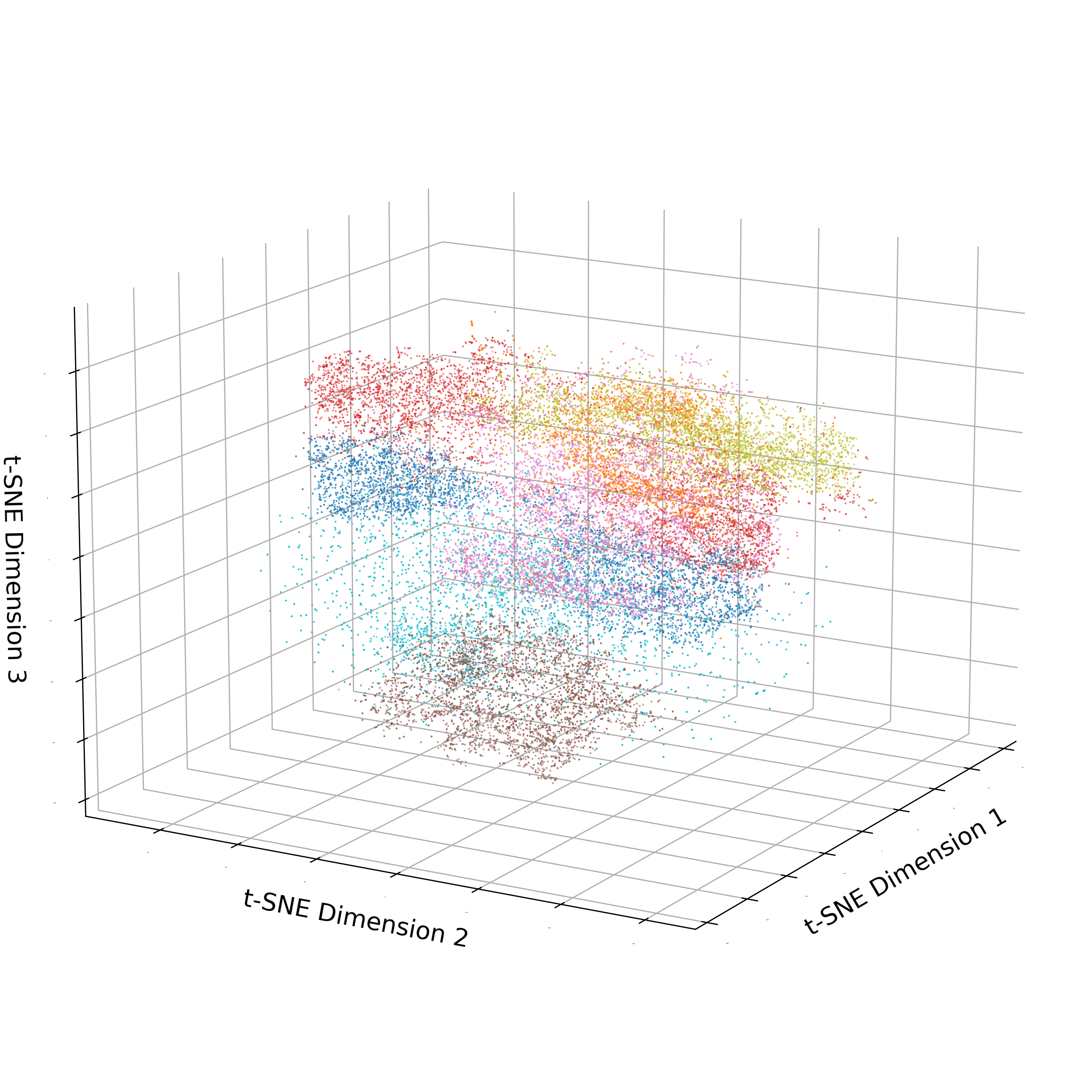}
\end{subfigure}\hfill
\begin{subfigure}{0.10\textwidth}
    \hspace*{-1cm}
    \raisebox{2cm}[0pt][0pt]{\includegraphics[width=2\linewidth,keepaspectratio]{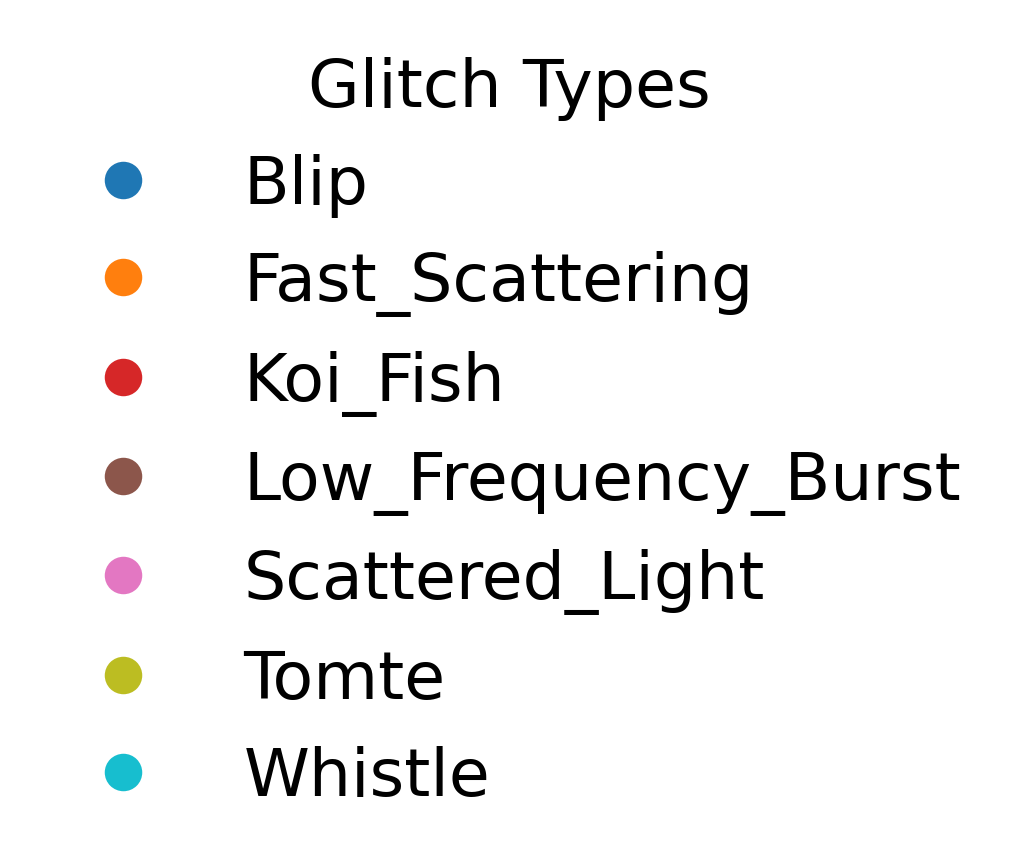}}
\end{subfigure}\hfill
\begin{subfigure}{0.45\textwidth}
    \hspace*{1cm}
    \includegraphics[width=\linewidth,height=0.26\textheight,keepaspectratio]{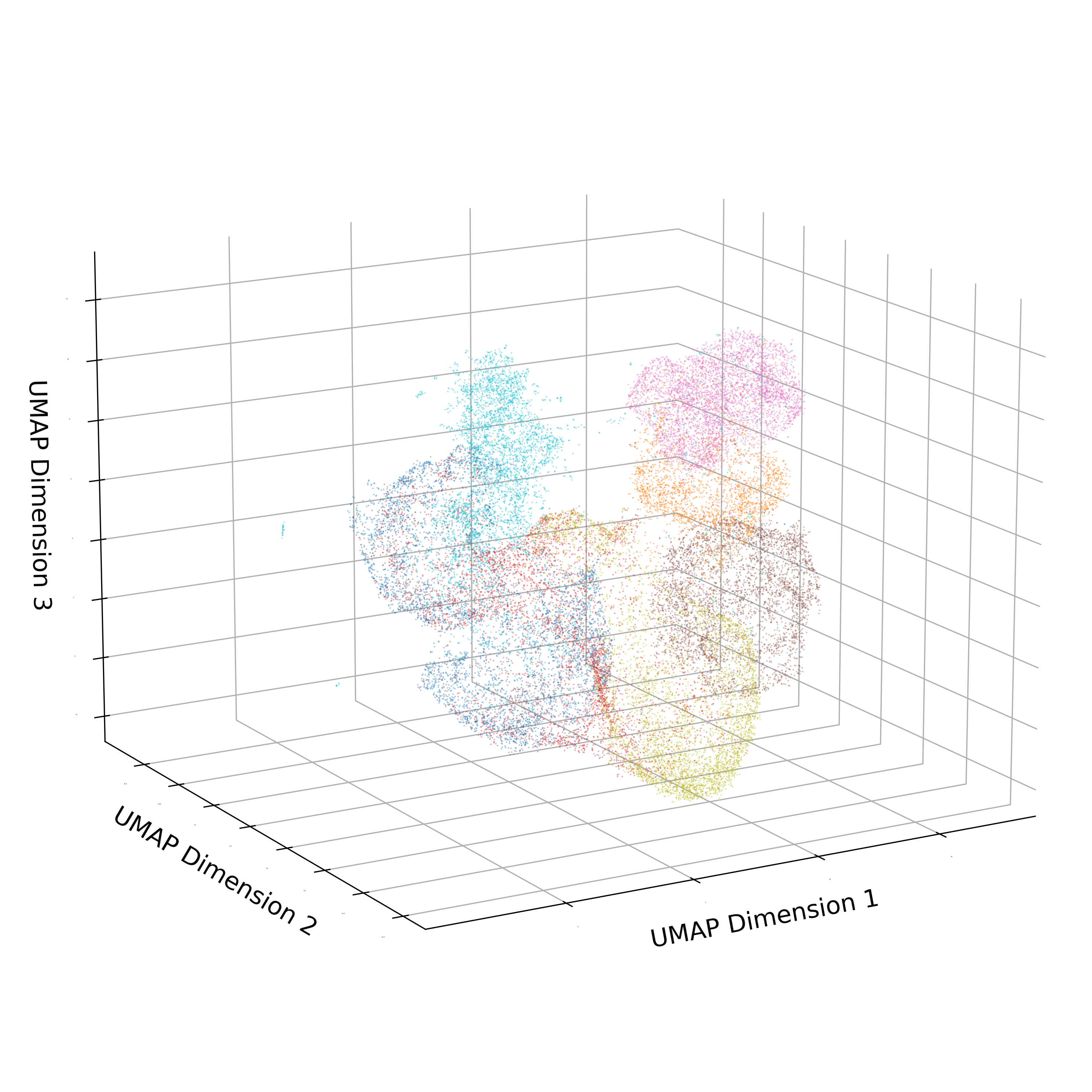}
\end{subfigure}
\vspace{-1.5em}
\begin{subfigure}{0.45\textwidth}
    \centering
    \includegraphics[width=\linewidth,height=0.26\textheight,keepaspectratio]{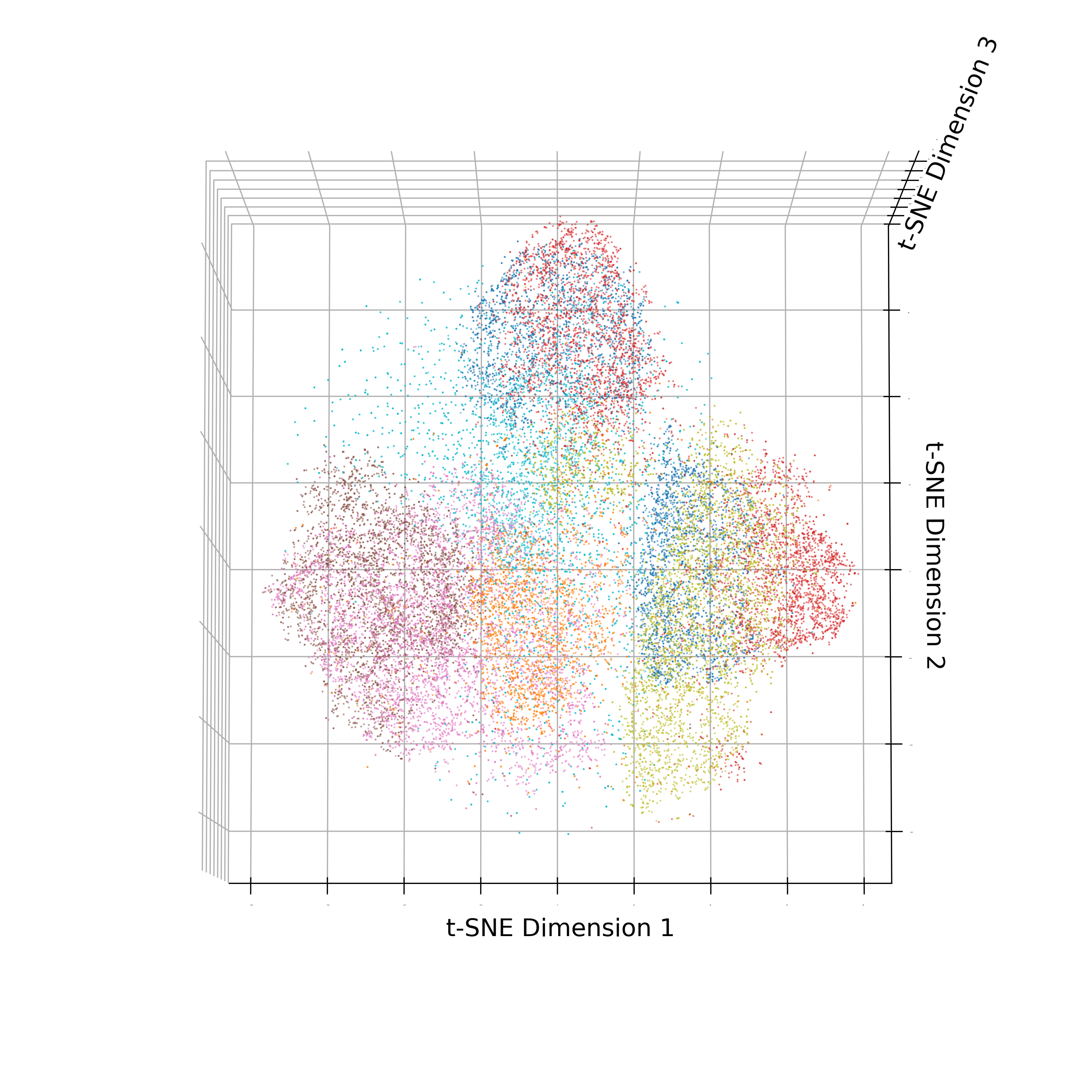}
\end{subfigure}\hfill
\begin{subfigure}{0.45\textwidth}
    \centering
    \includegraphics[width=\linewidth,height=0.26\textheight,keepaspectratio]{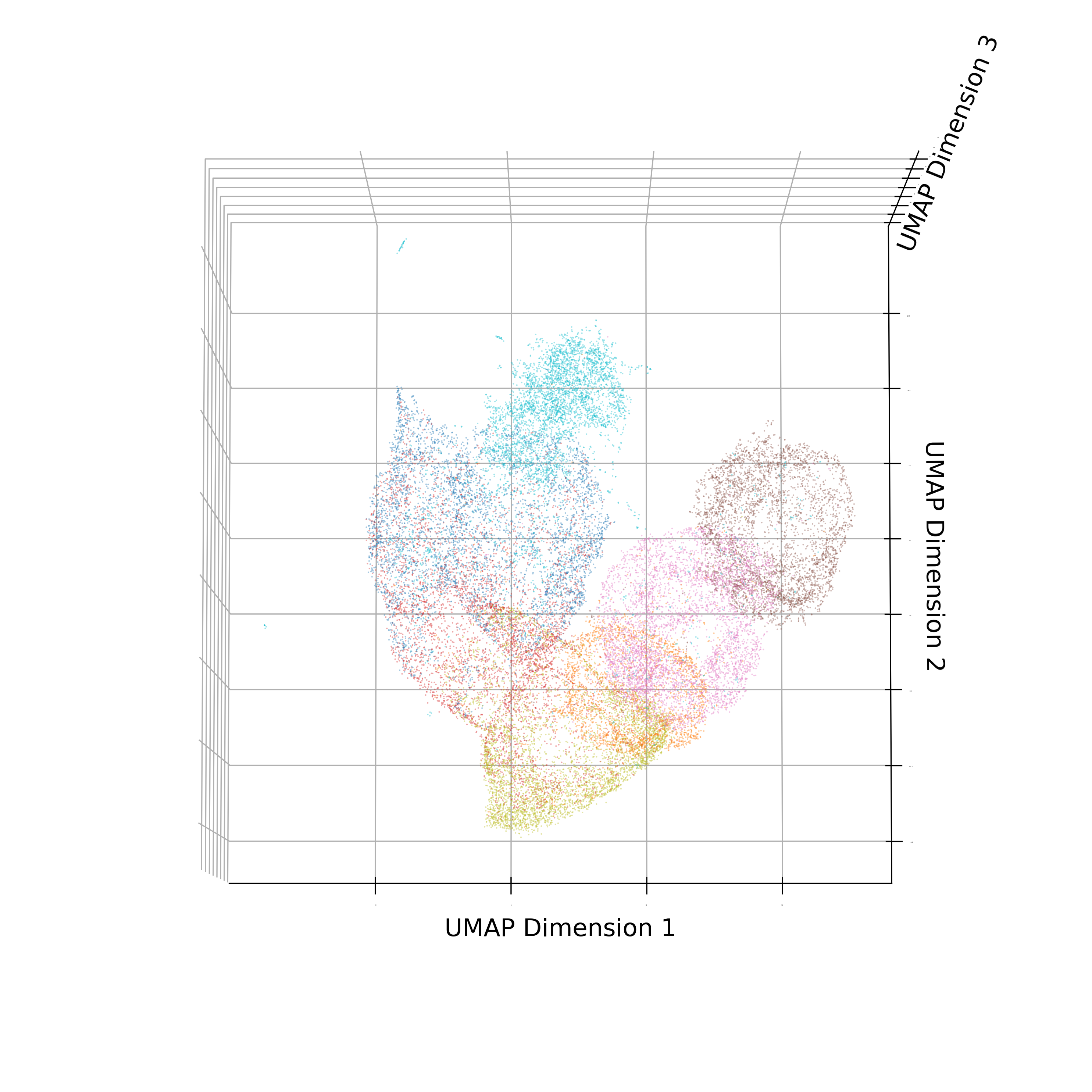}
\end{subfigure}
\caption{Three-dimensional t-SNE (left) and UMAP (right) embeddings of DeepExtractor reconstructions across three viewpoints. Colours correspond to Gravity Spy glitch classes. Well-separated clusters demonstrate that reconstructions preserve class-specific morphological structure in the time domain. The partial overlap between Blip and Koi Fish (visible in the middle row) is consistent with their known morphological proximity.}
\label{fig:de_3d_embeddings}
\end{figure*}

Figure~\ref{fig:de_3d_embeddings} shows three-dimensional t-SNE (left) and UMAP (right) embeddings, fitted on all 31,511 available DeepExtractor reconstructions, visualized from three complementary viewpoints.
The reconstructions form well-separated clusters that align closely with the Gravity Spy labels, demonstrating that morphologically distinct glitch types occupy distinct regions of the time-domain latent space.
Partial overlap between the Blip and Koi Fish clusters is visible, consistent with prior observations that these classes may share a physical origin~\citep{Areeda2017}.
The good clustering structure validates that the 31,511 DeepExtractor reconstructions used to train GlitchGAN are morphologically representative of seven distinct classes and are not dominated by reconstruction artifacts or label noise.

\subsection{Validation of GlitchGAN synthetic glitches}

\subsubsection{Classification evaluation using Gravity Spy}
\label{sec:GlitchGAN_gspy_eval}
\begin{figure*}[t]
\centering
\includegraphics[width=0.8\linewidth]{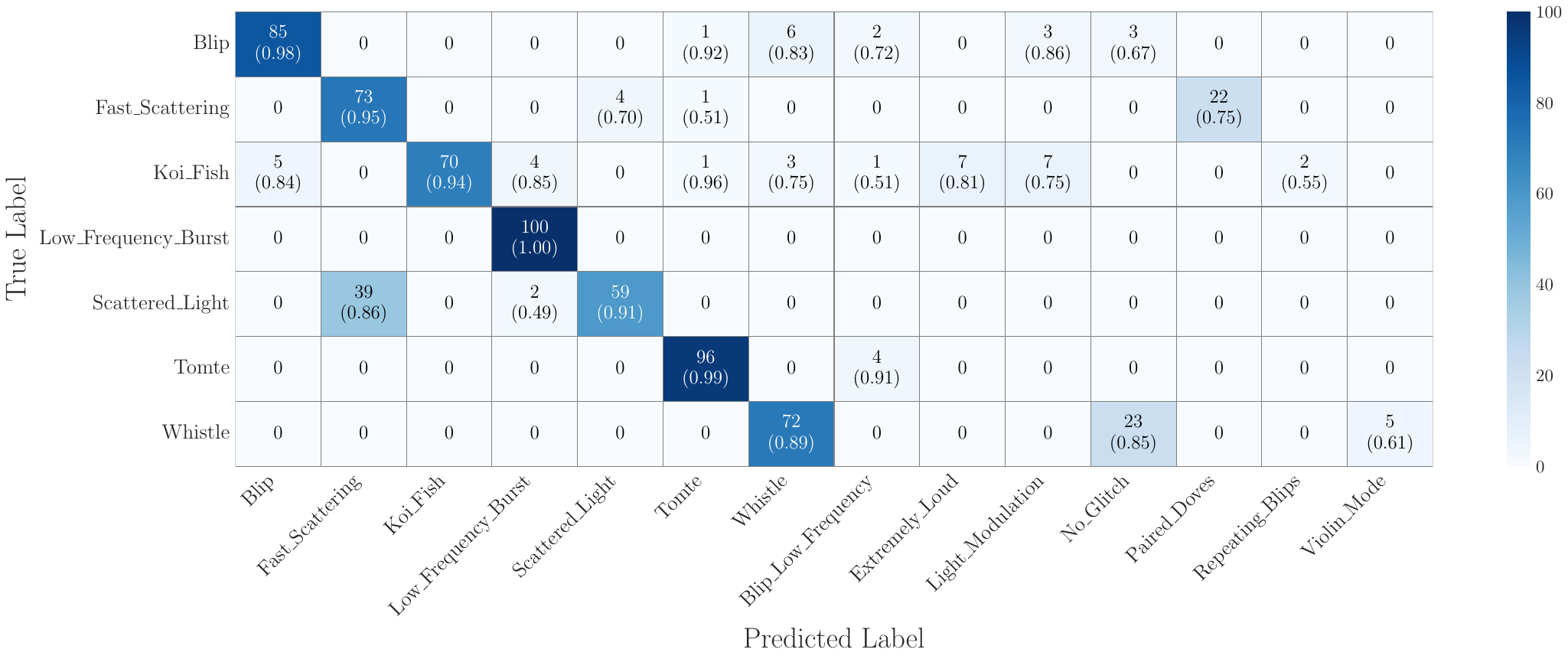}
\caption[Gravity Spy classification of synthetic GlitchGAN glitches]{
Confusion matrix showing Gravity Spy classifications of glitches generated by GlitchGAN, with 100 samples per class.
Each cell indicates the number of samples predicted for each class, with the average Gravity Spy confidence shown in brackets.}
\label{fig:cDVGAN_CM}
\end{figure*}

To quantify synthetic glitch realism we follow a similar approach to the DeepExtractor evaluation, injecting 100 generated glitches per class into quiet Hanford O3 background noise and classifying them with Gravity Spy.
The resulting confusion matrix shown in Figure~\ref{fig:cDVGAN_CM} reflects classifier predictions and confidence scores in brackets, with the full per-class breakdown summarized in Table~\ref{tab:gspy_ci}. 
The overall classification accuracy is $79.3\%$ (95\% Wilson CI [76.1, 82.1]\%).
While this is lower than the results for DeepExtractor reconstructions, the misclassifications are consistent with known limitations of Gravity Spy and the use of a fixed injected SNR according to the mean SNR of each glitch class.

Most misclassifications occur in the Koi Fish and Scattered Light classes. 
While only 70\% of the generated Koi Fish samples were correctly classified, the majority of the remaining samples were misidentified as Extremely Loud, Light Modulation and Blip.
This behaviour is consistent with known morphological similarities between Koi Fish and these glitch classes, which may arise from related physical mechanisms differing primarily in SNR~\cite{Areeda2017}.
Moreover, changing the injected SNR of the generated Koi Fish samples to 150 raises the classification accuracy to 79\% with an average confidence of $\sim$95\%, further illustrating the dependence of Gravity Spy classifications on the injected SNR.


This behaviour is further supported by injecting glitches at an SNR of 20.
In this lower-SNR regime, none of the generated Koi Fish samples were correctly classified.
Instead, they were misidentified as other morphologically similar classes.
A comparable SNR dependence is observed for the Whistle class, where 72\% of glitches were correctly identified.
Injecting the generated Whistle glitches at an SNR of 50 yields an accuracy of 94\%, whereas injecting them at an SNR of 20 leads to frequent misclassification as No Glitch.


For Scattered Light, Gravity Spy often predicts Fast Scattering, which is expected given their related scattering origins at low frequency. 
This was also observed for DeepExtractor's original reconstructions of both Fast Scattering and Scattered Light shown in Figure~\ref{fig:de_confusion}.

It is also important to note that some misclassifications may arise from stochastic interactions between the injected glitches and the specific noise realization into which they are injected, which can differ from the original conditions of the glitch sample. Additionally, labelling uncertainties in the Gravity Spy training set may contribute, and the discarded phase information in $Q$-scans may play a role—an issue we discuss further below.

Finally, we report that GlitchGAN generates 1000 glitches in approximately $22$ seconds on a CPU, demonstrating the model's efficiency for large-scale glitch synthesis.



\begin{table}[h]
\centering
\caption{Gravity Spy classification results for GlitchGAN-generated glitches ($n=100$ samples per class, 700 total). Accuracy intervals use the Wilson score method; confidence intervals use a percentile bootstrap (10{,}000 resamples).}
\label{tab:gspy_ci}
\resizebox{\linewidth}{!}{
\begin{tabular}{lcccc}
\toprule
\textbf{Class} & \textbf{Accuracy} & \textbf{95\% Wilson CI} & \textbf{Mean conf.} & \textbf{95\% bootstrap CI} \\
\midrule
Blip                 & 85.0\%  & [76.7, 90.7]  & 0.976 & [0.955, 0.993] \\
Fast Scattering      & 73.0\%  & [63.6, 80.7]  & 0.945 & [0.917, 0.970] \\
Koi Fish             & 70.0\%  & [60.4, 78.1]  & 0.939 & [0.915, 0.961] \\
Low Frequency Burst  & 100.0\% & [96.3, 100.0] & 1.000 & [0.999, 1.000] \\
Scattered Light      & 59.0\%  & [49.2, 68.1]  & 0.910 & [0.879, 0.939] \\
Tomte                & 96.0\%  & [90.2, 98.4]  & 0.992 & [0.981, 1.000] \\
Whistle              & 72.0\%  & [62.5, 79.9]  & 0.892 & [0.851, 0.930] \\
\midrule
Overall (700)        & 79.3\%  & [76.1, 82.1]  & 0.957 & [0.948, 0.965] \\
\bottomrule
\end{tabular}
}
\end{table}


\subsubsection{Global morphological similarity via UMAP}
\label{sec:glitchgan_umap}
\begin{figure*}[t]
\centering
\includegraphics[width=\linewidth]{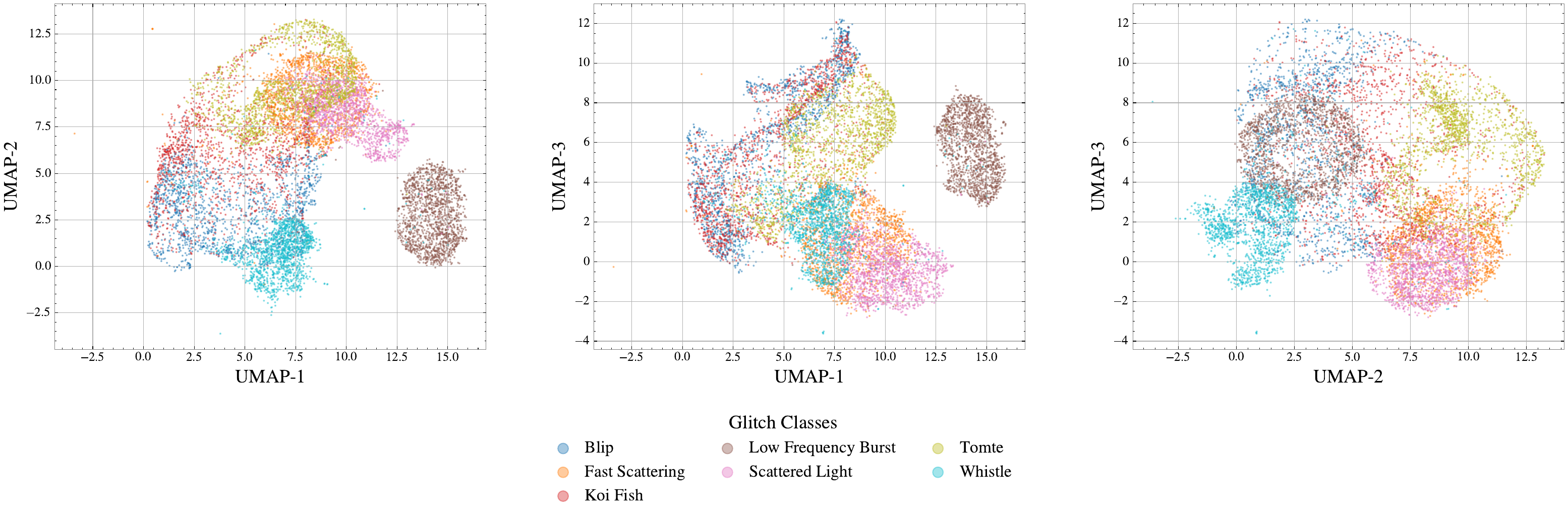}
\caption[Three perspectives of 2D UMAP embeddings comparing real and generated glitches]{
Three complementary views of two-dimensional UMAP embeddings showing both real and generated glitch samples.
Each colour corresponds to one of the seven glitch classes used during training. The embedding itself is fitted using 2,000 real and 2,000 generated samples per class (28,000 total). For visual clarity, this figure displays a randomly selected subset of 800 real and 800 generated samples per class (11,200 total).
The corresponding real vs synthetic class-wise UMAP projections using the same embedding are shown in Figure~\protect\ref{fig:umap_per_class_rotated}.}
\label{fig:umap_two_views}
\end{figure*}

To assess whether GlitchGAN captures the overall structure of the real glitch distribution, we project both real and synthetic glitch samples into a three-dimensional latent space using UMAP \citep{McInnes_2018}.
The embedding is fitted jointly to real and synthetic samples, using 2,000 real glitches per class (14,000 total) and 2,000 synthetic glitches per class (14,000 total), and the resulting embedding is visualized from three complementary perspectives in Figure~\ref{fig:umap_two_views}, showing two UMAP dimensions in each plot.


Figure~\ref{fig:umap_two_views} shows 800 real and 800 generated samples per class (11,200 total), randomly selected from the 28,000 samples used to fit the UMAP embedding.

\begin{figure*}[t]
\centering
\includegraphics[width=\linewidth]{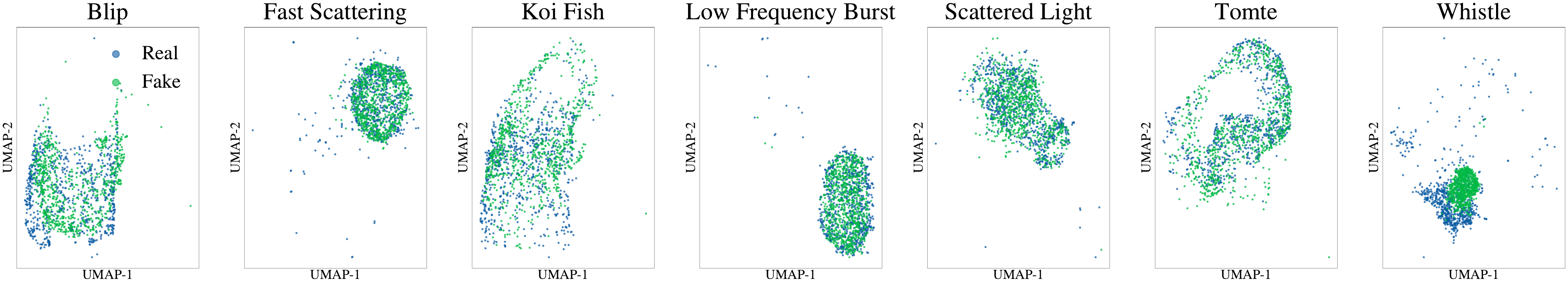}
\caption[Per-class UMAP embeddings of real and synthetic glitches]{
Two-dimensional UMAP embeddings for each glitch class, comparing real (blue) and GlitchGAN-generated (green) samples for the same UMAP embedding shown in Figure~\protect\ref{fig:umap_two_views}.
Each panel corresponds to one glitch type, showing the same 800 real and 800 generated samples per class plotted in Figure~\protect\ref{fig:umap_two_views}.
The close overlap between real and synthetic distributions across most classes demonstrates that GlitchGAN effectively learns class-specific morphological features, though some variation remains, particularly for the Whistle class.}
\label{fig:umap_per_class_rotated}
\end{figure*}

Corresponding individual class-wise UMAP projections using the same embedding are shown in Figure~\ref{fig:umap_per_class_rotated}, and provide detail to compare real vs. synthetic samples for each class.
Strong overlap between real (blue) and synthetic (green) samples indicates that GlitchGAN has learned the general morphological structure of the different glitch types.
The experiment embedding was fitted using different combinations of seeds and hyperparameters to ensure the patterns are consistent and not an artifact of one particular embedding. 
Similar patterns were observed across all iterations, confirming that the results are robust to the choice of hyperparameters and random seed.
The different configurations are described in Appendix~\ref{sec:tsne_umap_params}, and the full set of results for all nine iterations is available in the \texttt{evaluation\_plots/umap\_stability/} directory of our online repository (\url{https://github.com/tomdooney95/glitchgan}).

For most classes (Blip, Fast Scattering, Scattered Light, Tomte), synthetic and real samples show close clustering with substantial overlap.
The Whistle class exhibits the most noticeable divergence, with synthetic samples forming a partially separate cluster, suggesting that GlitchGAN captures the dominant Whistle morphology but does not fully explore the full morphological diversity within this class.
This may reflect limitations of the fixed GAN architecture when applied to this particular class without hyperparameter tuning, which is discussed in more detail below.


\subsubsection{Downstream classification of real vs. synthetic glitches\label{sec:downstream_classification_results}}

The CNN classifier distinguishes real from generated glitches with high a accuracy of 95.7\% (95\% CI [94.4, 96.7]\%).
A linear logistic regression baseline model fit directly on the raw waveform, with its L2 regularization strength of 100 selected via the validation split, is less capable of this discrimination, achieving 73.6\% accuracy, substantially lower than the corresponding CNN accuracy, although still well above random chance.
This mirrors a broader finding in the generative modelling literature: purpose-built classifiers routinely detect GAN-generated content with high accuracy even when it is otherwise realistic by coarser measures~\cite{wang2020cnngenerated}. 
Given the results of the previous experiments, we interpret this result not as evidence that GlitchGAN's synthetic glitches are morphologically dissimilar from real ones, but as a caveat on the resolution at which consistency holds against DeepExtractor's reconstructions.
For example, small artifacts in the generated glitches may be sufficient for a CNN to distinguish real from synthetic glitches even when they are otherwise morphologically similar. 
GlitchGAN reproduces class-level morphology and gross structural similarity faithfully, but synthetic glitches remain distinguishable from real ones under sufficiently targeted, fine-grained scrutiny.


Repeating this test with real and generated glitches injected into realistic detector noise reduces discriminability substantially. The CNN's accuracy drops to 66.0\% (95\% CI [63.3, 68.5]\%), while the linear baseline's accuracy falls to 58.9\% (95\% CI [56.2, 61.6]\%), with the L2 regularization strength of 0.0001 found during validation. This is consistent with the interpretation above: the CNN's advantage relied substantially on a fairly specific, consistent signature that is far less reliable once real and generated glitches are injected into realistic noise at per-sample SNRs drawn from the true empirical distribution (Section~\ref{sec:noise_injection_methodology}), rather than presented directly in the clean, noise-free representation used above. Under conditions that more closely resemble actual detector data (where synthetic glitches are expected to be used), physical consistency between real and synthetic glitches therefore holds more strongly than the clean-domain result alone would suggest.



\subsubsection{Multi-class classification and data augmentation}
\label{sec:multiclass_results}

Table~\ref{tab:fewshot_augmentation} shows seven-class classification accuracy, with real and generated glitches injected into realistic detector noise, across three real-data scarcity levels: abundant (2{,}206 real samples total, the full held-out training pool at its natural 70/10/20 split, $\sim$194--349 per class), moderately scarce (316 real samples total, 28--50 per class), and severely scarce (95 real samples total, 8--15 per class): Within each class two augmentation-target scales are shown: a modest target matching each pool's own majority-class count, and an aggressive target of 5{,}000 samples per class.

\begin{table*}[t]
\centering
\caption{Seven-class classification accuracy under realistic detector-noise injection, across three real-data scarcity levels and two augmentation-target scales: a modest target matching each training pool's own majority-class count, and an aggressive target of 5{,}000 samples per class. \textbf{Bold} entries are significantly better, and entries marked $^\dagger$ significantly worse, than that scarcity level's unaugmented baseline (non-overlapping 95\% Wilson CI).}
\label{tab:fewshot_augmentation}
\begin{tabular}{llcc}
\toprule
Scarcity level & Training set & Accuracy & 95\% CI \\
\midrule
\multirow{7}{*}{Abundant (70/10/20 split, $n=2206$, $\sim$194--349/class)}
 & Unaugmented                                          & 90.6\% & [88.1, 92.7]\% \\
 & Bootstrap-duplicated @ majority ($\sim$349/class)     & 89.8\% & [87.2, 92.0]\% \\
 & Augmented @ majority                                 & 91.4\% & [89.0, 93.4]\% \\
 & Synthetic-only @ majority                             & 89.0\% & [86.3, 91.2]\% \\
 & Bootstrap-duplicated @ 5000/class                     & 92.2\% & [89.8, 94.1]\% \\
 & Augmented @ 5000/class                                & 92.4\% & [90.0, 94.2]\% \\
 & Synthetic-only @ 5000/class                           & 90.9\% & [88.4, 92.9]\% \\
\midrule
\multirow{7}{*}{Moderately scarce (10/20/70 split, $n=316$, 28--50/class)}
 & Unaugmented                                          & 90.5\% & [89.2, 91.7]\% \\
 & Bootstrap-duplicated @ majority ($\sim$50/class)      & 89.6\% & [88.2, 90.8]\% \\
 & Augmented @ majority                                 & 90.5\% & [89.2, 91.7]\% \\
 & Synthetic-only @ majority                             & 89.3\% & [87.9, 90.5]\% \\
 & Bootstrap-duplicated @ 5000/class                     & 86.9\%$^\dagger$ & [85.5, 88.3]\% \\
 & Augmented @ 5000/class                                & 90.6\% & [89.3, 91.8]\% \\
 & Synthetic-only @ 5000/class                           & 90.7\% & [89.4, 91.8]\% \\
\midrule
\multirow{7}{*}{Severely scarce (3/20/77 split, $n=95$, 8--15/class)}
 & Unaugmented                                          & 85.7\% & [84.2, 87.0]\% \\
 & Bootstrap-duplicated @ majority ($\sim$15/class)      & 85.1\% & [83.6, 86.5]\% \\
 & Augmented @ majority                                 & 86.5\% & [85.1, 87.8]\% \\
 & Synthetic-only @ majority                             & 80.2\%$^\dagger$ & [78.6, 81.8]\% \\
 & Bootstrap-duplicated @ 5000/class                     & 87.4\% & [86.0, 88.6]\% \\
 & Augmented @ 5000/class                                & \textbf{91.0\%} & [89.8, 92.0]\% \\
 & Synthetic-only @ 5000/class                           & \textbf{91.8\%} & [90.6, 92.8]\% \\
\bottomrule
\end{tabular}
\end{table*}

At the data-abundant level, the full held-out real training pool at its natural 70/10/20 split (2{,}206 real samples total, $\sim$194--349 per class) -- augmentation provides at most a modest benefit. At the majority-class target ($\sim$349/class), none of bootstrap-duplication (89.8\%), augmentation (91.4\%), or synthetic-only training (89.0\%) differ significantly from the unaugmented baseline (90.6\%). Scaled up to 5{,}000 samples per class, bootstrap-duplication (92.2\%), augmentation (92.4\%), and synthetic-only training (90.9\%) are all nominally higher than baseline, though none reach statistical significance at this test-set size.

To assess whether this changes as real data becomes scarcer, we repeat the experiment with a moderately scarce real training pool (316 samples total, 28--50 per class; middle third of Table~\ref{tab:fewshot_augmentation}). At the majority-class target ($\sim$50/class), none of bootstrap-duplication (89.6\%), augmentation (90.5\%), or synthetic-only training (89.3\%) differ significantly from the unaugmented baseline (90.5\%), consistent with the classification task already being close to its accuracy ceiling with this much real data available. Scaled up to 5{,}000 samples per class, a different pattern emerges: bootstrap-duplication becomes significantly \emph{worse} than the unaugmented baseline (86.9\% vs.\ 90.5\%), while GlitchGAN-augmented and synthetic-only training remain statistically indistinguishable from it (90.6\% and 90.7\%, respectively). This harmful-duplication effect was absent at the data-abundant scale above, where the duplication factor was much milder ($\sim$18$\times$ versus roughly $100$--$180\times$ here), suggesting the harm from naive duplication is tied to how extreme the duplication factor is, rather than being a general property of oversampling real data to a large nominal training set size. Even so, synthetic augmentation itself remains a \emph{safe} choice in this regime, providing no measurable benefit but no harm either, unlike naive duplication.

Finally, at the most severely scarce end of this sweep, the picture depends strongly on target scale. At the modest majority-class target ($\sim$15/class), bootstrap-duplication (85.1\%) and augmentation (86.5\%) are statistically indistinguishable from the unaugmented baseline (85.7\%), while training on synthetic data alone is significantly \emph{worse} (80.2\%, non-overlapping 95\% CI) -- unsurprising, since this configuration discards the handful of real anchors entirely in favour of an equally small, purely synthetic set. Scaling the same three strategies up to 5{,}000 samples per class inverts this pattern: augmentation (91.0\%) and synthetic-only training (91.8\%) both become significantly better than baseline (non-overlapping 95\% CIs), while bootstrap-duplication (87.4\%) still shows no significant improvement. Since the augmented and duplicated configurations at this scale are matched in total training set size, this isolates the source of the improvement: it is GlitchGAN's genuine morphological diversity, not merely a larger nominal training set, that drives the gain -- and synthetic augmentation only pays off once deployed at sufficient scale.

Together, these results demonstrate a practical use case for GlitchGAN: in data-scarce scenarios, synthetic augmentation can improve downstream classifier performance, provided it is deployed at sufficient scale. As real data becomes more abundant, this benefit diminishes and eventually vanishes entirely, consistent with the classification task saturating quickly regardless of data source. Naive duplication's harm at large nominal scale is itself scarcity-dependent: it appears only when the real pool is small enough that reaching a large training set requires an extreme duplication factor.
Nevertheless, GlitchGAN augmentation is never harmful, showing that it effectively learned the morphological structure of the real glitch distribution.

\subsubsection{Generating hybrid glitches via class interpolation}

A key advantage of class-conditional generative models is the ability to produce ``hybrid'' or transitional morphologies by sampling mixed class vectors.

We explore two sampling strategies:

\begin{enumerate}
\item
\textbf{Simplex mixing:} The class vector is normalized such that all coefficients sum to 1, i.e., sampled from a $k=7$ simplex. This constrains mixing proportions to sum to 100\%.

\item
\textbf{Uniform mixing:} Each class coefficient is independently sampled from $[0, 1]$ without normalization constraint, allowing more extreme interpolations.
\end{enumerate}

\begin{figure*}[t]
\centering
\begin{subfigure}[t]{0.48\linewidth}
    \centering
    \includegraphics[width=\linewidth]{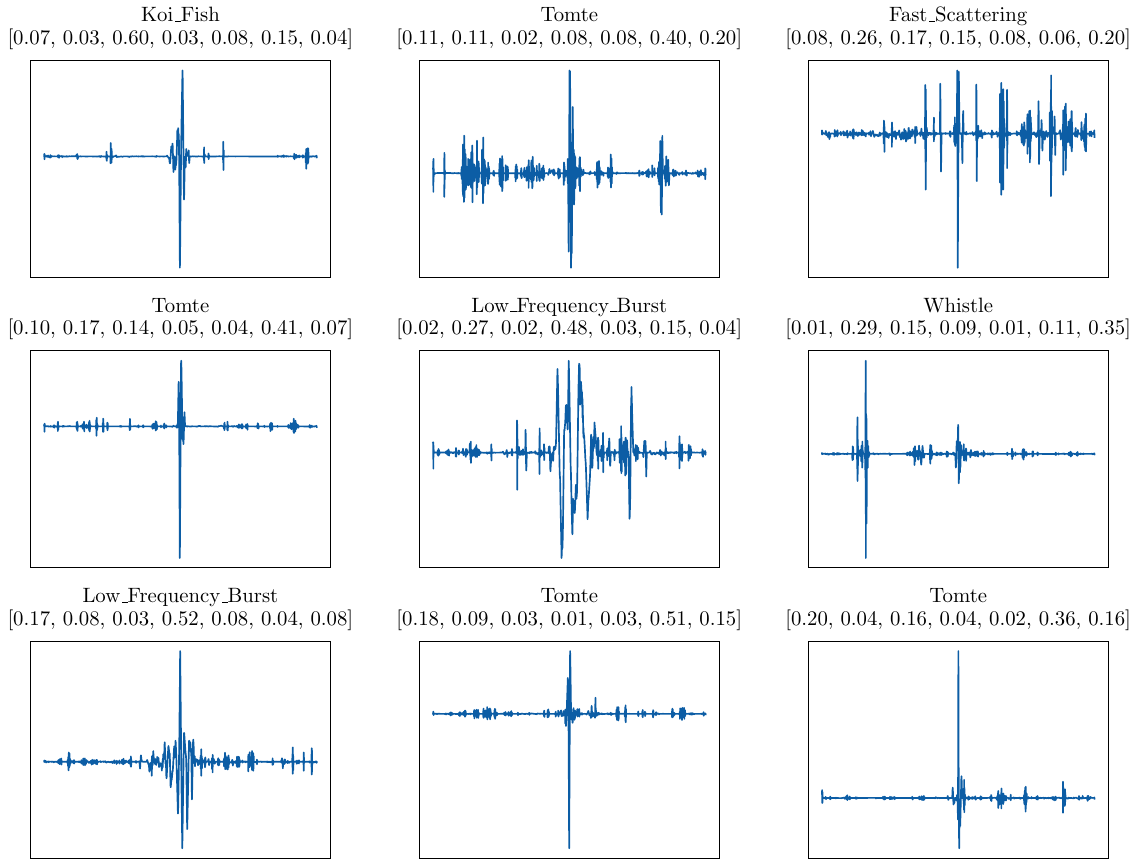}
    \caption{Simplex mixing ($\sum c_i = 1$).}
    \label{fig:mixing_simplex}
\end{subfigure}
\hfill
\begin{subfigure}[t]{0.48\linewidth}
    \centering
    \includegraphics[width=\linewidth]{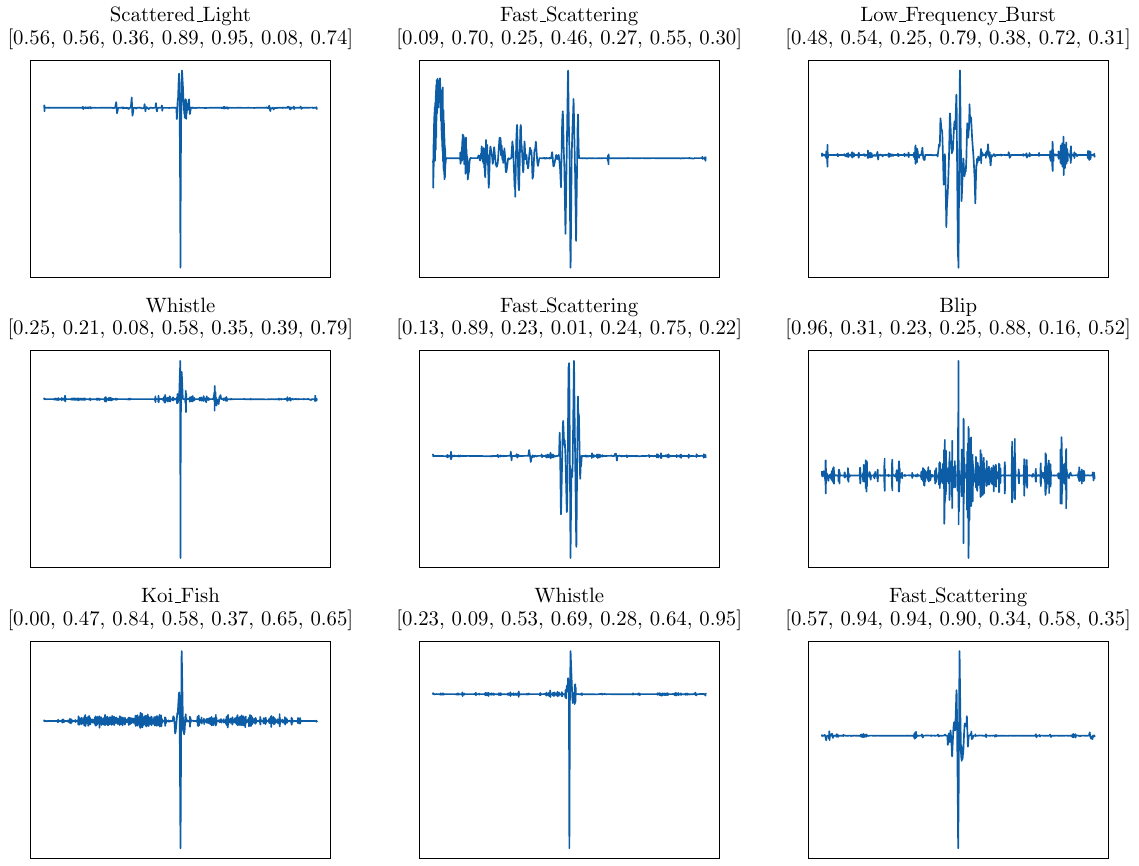}
    \caption{Uniform mixing ($c_i \in [0,1]$).}
    \label{fig:mixing_uniform}
\end{subfigure}

\caption[Hybrid glitch generation via class-conditioned mixing]{
Examples of hybrid glitch waveforms generated by GlitchGAN using two class-conditioning strategies.
\textbf{(a)} Simplex mixing samples each class vector from a $k=7$ simplex, ensuring coefficients sum to one.
\textbf{(b)} Uniform mixing samples each coefficient independently from a uniform distribution in U$[0, 1]$.
The corresponding sampled class vectors are shown above each plot, with the dominant glitch label also shown. Both methods produce hybrid morphologies interpolating between glitch classes, demonstrating GlitchGAN's ability to explore smooth transitions across the learned latent class space.}
\label{fig:hybrid_glitch_mixing}
\end{figure*}

\begin{figure}[ht]
\centering
\includegraphics[width=\linewidth]{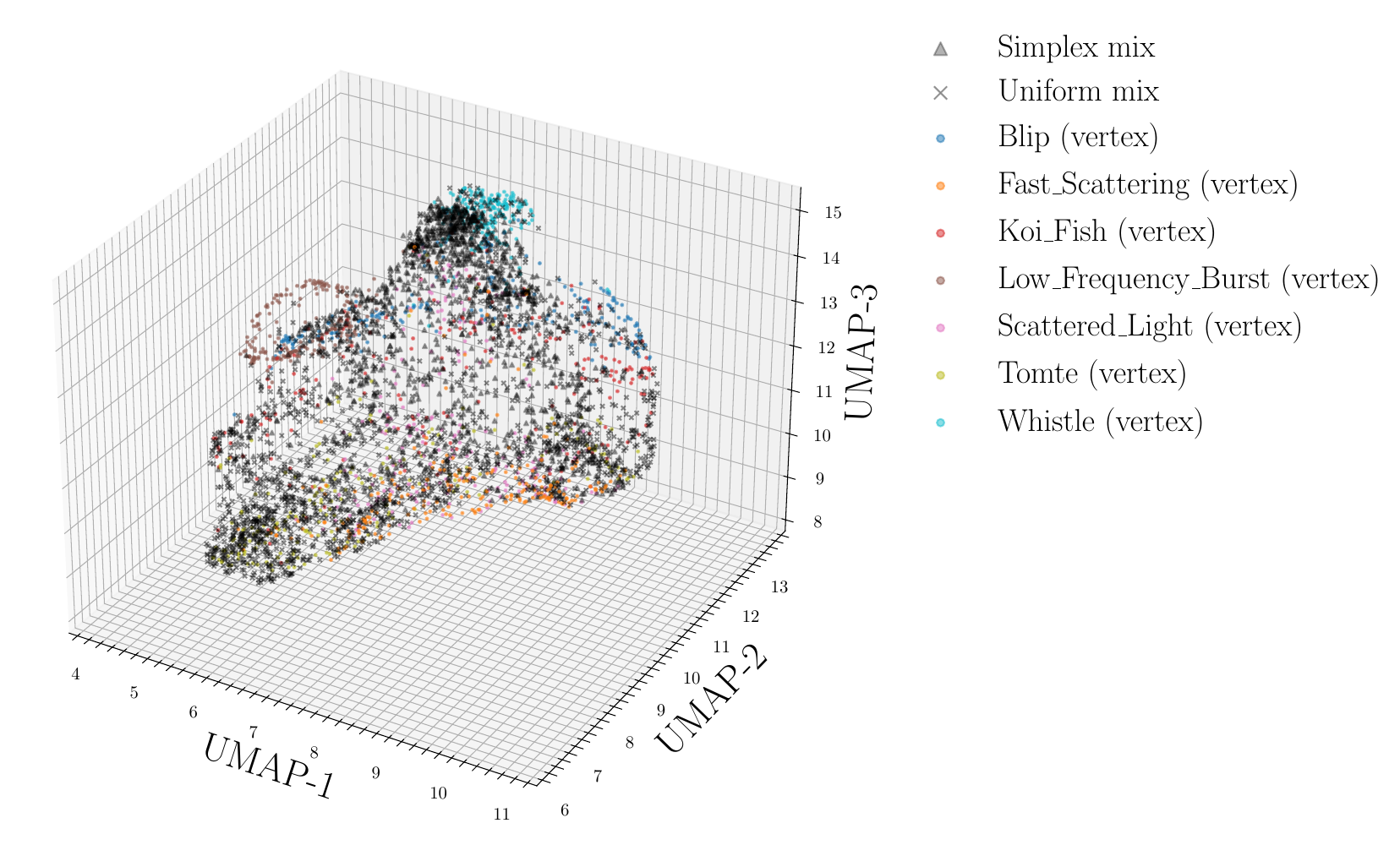}
\caption[UMAP embedding of class-mixed and vertex glitch samples]{
Three-dimensional UMAP embedding showing the distribution of class-mixed glitch samples relative to the vertex (pure-class) glitches.
Vertex samples are coloured by their respective glitch class, while class-mixed samples generated using simplex and uniform class-conditioning vectors are shown in black.
The smooth interpolation of mixed samples between class clusters illustrates how GlitchGAN captures continuous transitions across glitch morphologies in the latent class space.}
\label{fig:simplex_uniform_umap}
\end{figure}

Figure~\ref{fig:hybrid_glitch_mixing} shows glitch examples from both mixing strategies, demonstrating intermediate morphologies.
As shown, samples from these datasets exhibit mixed morphological characteristics drawn from multiple glitch classes as per the sampled class vector.
The UMAP projection of mixed samples in Figure~\ref{fig:simplex_uniform_umap} confirms that these hybrid samples smoothly interpolate between pure-class point clouds, with uniform mixing exploring the broadest region of the latent space.

It is important to note that these hybrid glitches do not represent real transitional glitch mechanisms. 
Rather, they are synthetic class mixtures produced by interpolating the generative model's class-conditioning input, which the model maps to smoothly and continuously varying morphologies.
While hybrid samples are not further analyzed in this study, they have potential applications for pipeline validation and data augmentation for glitch detection or classification systems.

\subsection{Limitations of magnitude-only representations}

We further evaluated the DiT model as a comparison; the full Gravity Spy and UMAP results are shown in Appendix~\ref{app:dit_evaluation}.
DiT-generated samples overlap with real glitches primarily for low-frequency classes (e.g., Fast Scattering, Scattered Light), but show reduced diversity and separable clusters for most other classes where GlitchGAN was more successful.

\begin{figure*}[t]
\centering
\includegraphics[width=0.8\linewidth]{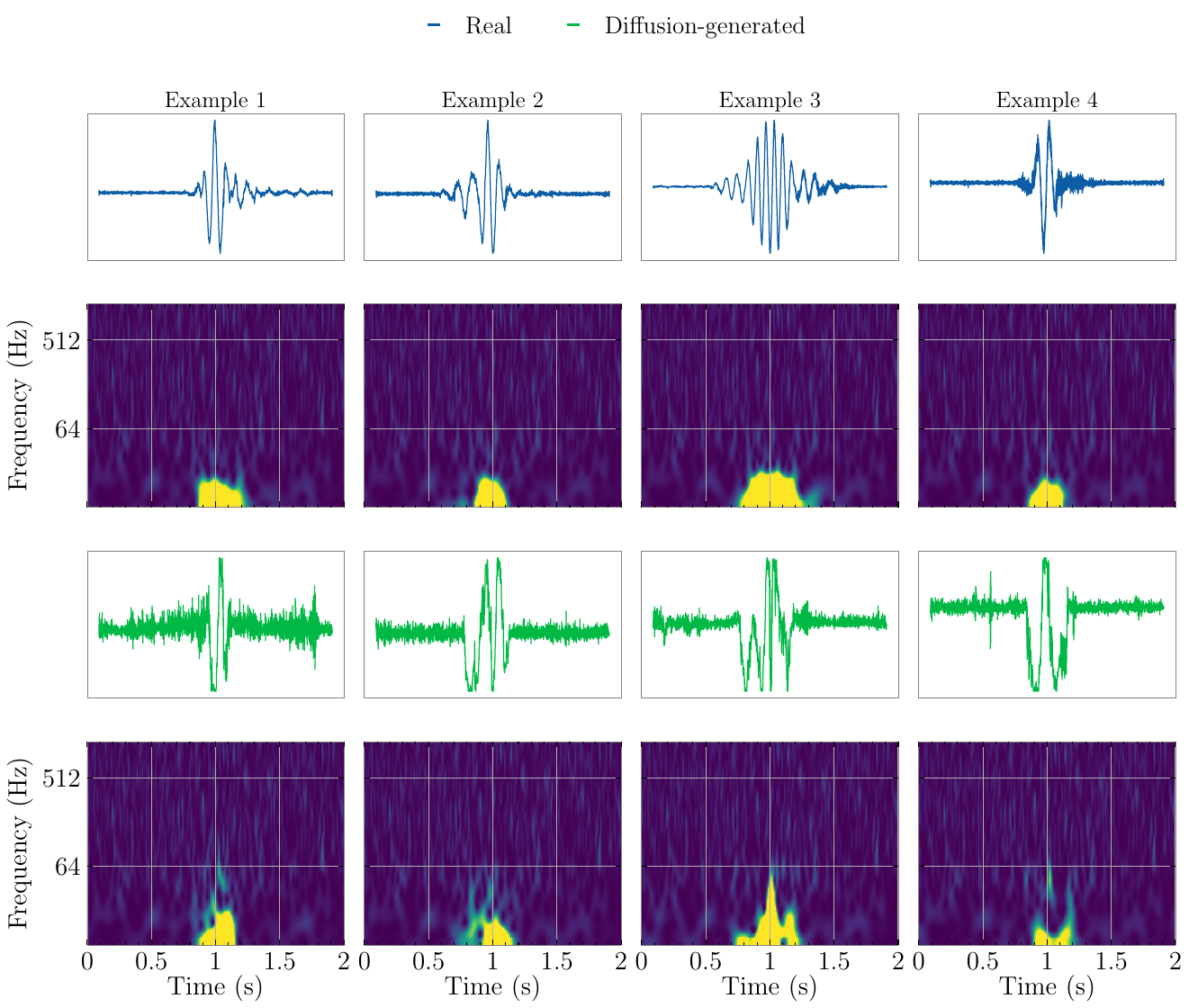}
\caption{Four examples each of real (top two rows) and DiT-generated (bottom two rows) Low Frequency Burst glitches, showing both the time-domain waveform and the magnitude $Q$-scan representation. The time-domain waveforms are clearly distinct, yet the $Q$-scans appear similar, explaining why Gravity Spy classifies the DiT samples correctly with high confidence.}
\label{fig:dit_lfb}
\end{figure*}

A notable example is the Low Frequency Burst class: UMAP reveals two clearly distinct clusters (real vs.\ synthetic), yet Gravity Spy classifies 87\% of DiT-generated samples correctly with 97\% mean confidence.
Figure~\ref{fig:dit_lfb} illustrates why: while the time-domain waveforms of real and DiT-generated Low Frequency Bursts are clearly distinct, their magnitude $Q$-scan representations appear similar, explaining the classifier's high confidence despite the physically unrealistic time-domain morphology.


This finding underscores a critical limitation: classifiers operating on magnitude-only representations cannot reliably assess glitch realism.
Complementary validation methods preserving phase information are essential for ensuring synthetic glitches are suitable for detector simulations.

Overall, while this specific DiT implementation demonstrates that latent diffusion is a viable approach to conditional time-domain glitch generation, its ability to reproduce full morphological diversity---especially for mid- and high-frequency content---is limited relative to GlitchGAN under the configuration tested here.
It is important to note that the DiT model implemented serves primarily to highlight the limitation in magnitude-only classification.
Better diffusion architectures and training strategies may improve performance beyond GlitchGAN and other GAN architectures.

\section{Discussion}
\label{sec:discussion}

\subsection{Performance assessment and limitations}

GlitchGAN successfully demonstrates time-domain synthesis of multiple glitch classes reconstructed by DeepExtractor.
Substantial UMAP overlap between real and synthetic samples, combined with accurate Gravity Spy classification of generated glitches indicates morphological fidelity of generated glitches.
Furthermore, the ability of GlitchGAN-generated samples to augment real training datasets for downstream classification demonstrates practical utility and shows the model is capable of learning relevant features of real glitch reconstructions.
However, there exists several limitations and caveats which we address here.

Gravity Spy classification and UMAP embeddings are comparatively coarse validation metrics, operating at the level of class-wide structure rather than fine-grained statistical detail. A dedicated real-vs-generated CNN classifier (Section~\ref{sec:downstream_classification_methods}) detects a residual statistical signature distinguishing synthetic from real glitches with high accuracy in the clean, noise-free domain, indicating that subtle, nonlinear differences remain even where these coarser methods show close agreement. This detectability is substantially, though not completely, reduced once glitches are considered under realistic detector-noise conditions, suggesting the residual signature is most apparent in idealized representations rather than in the conditions under which synthetic glitches would actually be deployed.

Partial failure was observed for the Whistle class, possibly due to insufficient regularization or architectural constraints on receptive field size. 
For example, smaller kernel sizes in generator and discriminator  layers may be needed to capture the high-frequency content of Whistle glitches (please see Appendix~\ref{sec:cDVGAN_Architecture} for details).

Since GlitchGAN is trained on normalized waveforms, it does not learn an explicit SNR distribution, and thus generated glitches must be rescaled to a specified SNR for use in downstream applications. 
As shown in Section~\ref{sec:GlitchGAN_gspy_eval}, Gravity Spy classifications are strongly sensitive to the injected SNR --- consistent with the fact that individual glitch classes are partly characterized by their SNR distribution.
Appropriate rescaling is therefore an important consideration for simulations and pipeline validation studies using these synthetic samples.
For example, one may choose to rescale generated glitches to match the SNR distribution of the real glitches within a given class, or to a fixed SNR for specific applications such as testing pipeline sensitivity at particular SNR thresholds.
To ensure fidelity for downstream applications, Gravity Spy can be used to verify that the rescaled glitches are classified as the intended class with high confidence.

Given the empirical relationships that exist between SNR and glitch morphology, at least for certain classes (e.g. Blip vs. Koi Fish \cite{Areeda2017}), it is important to consider the implications of rescaling generated glitches to a fixed SNR.
Normalizing glitches to a fixed SNR, or to the class-wide SNR distribution, may alter the morphology of the generated glitches relative to the real population, which is an important consideration for downstream applications.
A more realistic rescaling strategy could instead use a $k$-nearest-neighbours (kNN) approach, in which each generated glitch is assigned an SNR sampled from the $k$ real training glitches nearest to it in morphology space, thereby preserving the class-internal SNR-morphology correlations that a single fixed or class-averaged SNR would erase.
Jointly modelling glitch morphology and SNR is a more complete, though more complicated, direction for future work to achieve fully realistic synthetic glitches, which could be achieved by extending the class-conditional generator to include an additional SNR-conditioning input.

While DeepExtractor shows great potential for glitch reconstruction, it is not perfect and may introduce artifacts or distortions in the reconstructed glitches.
These artifacts may propagate into the GlitchGAN training data, potentially affecting the quality of generated glitches, which is difficult to quantify for real glitches since we lack ground truth glitch waveforms.
However, the strong agreement between real and synthetic glitches in UMAP embeddings and Gravity Spy classifications, and the capability of the synthetic data to effectively train CNNs to classify real glitch classes, suggest that any artifacts introduced by DeepExtractor are not dominant in the training data.
That being said, should a more accurate glitch reconstruction method become available in the future, it could be used to retrain GlitchGAN and potentially improve the quality of generated glitches.

Finally, we note that no hyperparameter optimization was performed for this specific dataset; the model is evaluated ``out-of-the-box'' from previous work (except for the dimensionality of the discriminator input, generator output and class vectors), representing a conservative assessment of GlitchGAN's potential.
Tuning the parameters of the generator and discriminator architectures, as well as the training procedure, could further improve performance, particularly for the more challenging Whistle class.








\subsection{Practical applications}

The synthetic glitch population produced by GlitchGAN is immediately useful for:

\begin{enumerate}
\item
\textbf{Mock data challenges (MDCs):} Enriching realistic detector simulations with synthetic instrumental noise

\item
\textbf{Detector characterization:} Validating glitch detection, mitigation, and subtraction algorithms

\item
\textbf{Training data augmentation:} Supplementing limited real glitch samples with GlitchGAN-generated samples, demonstrated in Section~\ref{sec:multiclass_results} to improve downstream classification accuracy in data-scarce regimes

\item
\textbf{Pipeline robustness testing:} Assessing pipeline sensitivity and false-alarm rates in the presence of instrumental noise morphologies
\end{enumerate}

\subsection{Recommendations for future work}

There are several promising directions for future research building on this work. Firstly, hyperparameter optimization and architectural refinements could be performed to further improve the quality of generated data, particularly for the more challenging Whistle class.
Architecture improvements may include the incorporation of attention mechanisms, smaller convolutional kernel sizes or dilated convolutions for multi-scale feature extraction.

Ultimately, the goal is to develop a comprehensive generative model capable of synthesizing the entire range of glitch morphologies in GW detectors with high fidelity, enabling robust simulations and pipeline validation for current and future GW detectors.
Future work will iteratively include more glitch classes, starting with more common and progressively incorporating less common ones.

It would be beneficial to develop more sophisticated validation metrics that explicitly evaluate phase consistency, not just magnitude-level morphology, to ensure synthetic glitches are suitable for detector simulations.
Such methods will be beneficial for evaluating the realism of synthetic glitches and guiding future improvements in generative modeling approaches.

While class conditional generation is a powerful tool, it may also be desirable to develop conditional generation strategies allowing control over continuos features such as glitch duration, SNR, and frequency content, independently from class morphology.
This would enable more flexible synthesis of glitches for specific applications, such as testing pipeline sensitivity to particular frequency bands or durations.

Finally, integrating these features with mock data challenge pipelines for rapid deployment in detector simulations will be an important step towards practical utility of these synthetic glitch populations for the GW data analysis community.







\section{Conclusions}

We have demonstrated that a class-conditional deep generative model called GlitchGAN can effectively synthesize realistic GW detector glitches directly in the time domain at speed, generating 1000 glitch samples in under 22 seconds on a CPU.
By leveraging high-fidelity reconstructions from a glitch reconstruction framework called DeepExtractor and training a single unified GlitchGAN model based on the cDVGAN architecture across seven glitch classes, we achieve synthetic samples with strong morphological agreement with real data.
GlitchGAN's synthetic glitch population offers practical utility for mock data challenges, detector characterization studies, and validation of detection pipelines.

We evaluated both DeepExtractor's reconstructions used to train GlitchGAN and GlitchGAN's synthetic data using two complementary approaches: a supervised classification analysis using Gravity Spy, a state-of-the-art glitch classification algorithm trained on an extensive labeled dataset, and an unsupervised, data-driven UMAP analysis to visualize and compare the class distributions of real and synthetic glitches in a projected three-dimensional space.

The Gravity Spy classification results for DeepExtractor reconstructions showed that 89.57\% of the reconstructed glitches were correctly classified, with a mean classifier confidence of 96.77\% for correct predictions.
Misclassified samples showed a lower confidence of $84.84\%$, suggesting that classifier confidence can serve as a quality veto.
Overall, the misclassification patterns are generally explainable, typically misclassifying as morphologically similar classes, and do not indicate systematic failure of the reconstruction.
The unsupervised UMAP analysis of DeepExtractor reconstructions revealed well-separated clusters according to the Gravity Spy labels, demonstrating that the reconstructions preserve class-specific morphological structure in the time domain.

For GlitchGAN's generated glitches, Gravity Spy classified most samples correctly with an overall accuracy of $~79\%$.
Misclassifications primarily occurred for the Koi Fish and Scattered Light classes, which can largely be attributed to morphological similarities between similar classes and Gravity Spy's classification being dependent on SNR.
For instance, Scattered Light glitches were often misclassified as Fast Scattering, while Koi Fish are often misclassified as Extremely Loud or Light Modulation.
These overlaps are expected, as the classes share similar time-frequency signatures.
This is confirmed by changing the injected SNR to 150 for Koi Fish samples and 50 for the Whistle class, where 79\% and 94\% are correctly classified, respectively.

The UMAP results showed that GlitchGAN-generated glitches generally occupy the same regions as real data in the embedded UMAP space, indicating that the model captures much of the morphological diversity within each glitch class.
One exception is the Whistle class, which consists of the highest-frequency features in the dataset.
For this class, synthetic samples formed a partially distinct cluster from the real data, suggesting that GlitchGAN did not fully capture the fine-scale, high-frequency structure characteristic of these glitches.
Future work could address this by incorporating smaller convolutional kernels in the network (balanced against the need for long-range temporal modelling), dilated convolutions to better capture hierarchical structure, or attention mechanisms to enhance global feature representation.

Beyond these coarser validation methods, we further tested synthetic glitch quality using a dedicated classifier trained specifically to discriminate real glitches from GlitchGAN-generated ones. This classifier reliably detects synthetic samples in a clean, noise-free representation, indicating that subtle statistical differences remain even where class-level and structural similarity are strong. However, this detectability narrows substantially once glitches are injected into realistic detector noise, suggesting that physical consistency between real and synthetic glitches holds more strongly under conditions resembling actual detector data where GlitchGAN glitches are intended for use.

Despite this residual detectability, we demonstrated a concrete practical use case for GlitchGAN: augmenting real training sets with synthetic samples matched simple duplication of the same real data when data was abundant, and increasingly outperformed it as real data became scarcer, remaining safe where duplication turned harmful and improving downstream multi-class glitch classification accuracy in the most severely data-scarce regime. 
This indicates that GlitchGAN's synthetic glitches carry genuine, useful morphological information beyond simply increasing dataset size.

While it is not investigated explicitly how reconstruction artifacts from DeepExtractor may affect GlitchGAN's performance, the strong agreement between real and synthetic glitches in UMAP embeddings and Gravity Spy classifications suggests that any artifacts introduced by DeepExtractor are not dominant in the training data, and GlitchGAN is effective at learning the distributions reconstructed by DeepExtractor.
Furthermore, should a more accurate glitch reconstruction method become available in the future, it could be used to retrain GlitchGAN and potentially improve the quality of generated glitches.

Finally, this study highlights a fundamental limitation in using purely magnitude-based $Q$-scan representations for supervised glitch classification.
We observed that a less robust diffusion-based model, which produces unrealistic Low Frequency Burst glitches in the time domain, can still yield correct Gravity Spy labels with high confidence.
Although the time-domain waveforms of real and synthetic samples are clearly distinct, their $Q$-scan representations appear similar, explaining the classifier’s high confidence.
This discrepancy is evident in the corresponding UMAP analysis of time-domain data, which reveals two distinct clusters separating real and synthetic samples.
This finding highlights the importance of preserving phase information and integrating unsupervised analyses to complement glitch characterization efforts.


\section*{Acknowledgements}


This research has made use of data or software obtained from the Gravitational Wave Open Science Center, a service of the LIGO Scientific Collaboration, the Virgo Collaboration, and KAGRA. This material is based upon work supported by NSF's LIGO Laboratory which is a major facility fully funded by the National Science Foundation, as well as the Science and Technology Facilities Council (STFC) of the United Kingdom, the Max-Planck-Society (MPS), and the State of Niedersachsen/Germany for support of the construction of Advanced LIGO and construction and operation of the GEO600 detector. Additional support for Advanced LIGO was provided by the Australian Research Council. Virgo is funded, through the European Gravitational Observatory (EGO), by the French Centre National de Recherche Scientifique (CNRS), the Italian Istituto Nazionale di Fisica Nucleare (INFN) and the Dutch Nikhef, with contributions by institutions from Belgium, Germany, Greece, Hungary, Ireland, Japan, Monaco, Poland, Portugal, Spain. KAGRA is supported by Ministry of Education, Culture, Sports, Science and Technology (MEXT), Japan Society for the Promotion of Science (JSPS) in Japan; National Research Foundation (NRF) and Ministry of Science and ICT (MSIT) in Korea; Academia Sinica (AS) and National Science and Technology Council (NSTC) in Taiwan.
The authors are grateful for computational resources provided by the LIGO Laboratory and supported by the National Science Foundation Grants No. PHY-0757058 and No. PHY-0823459.

\vspace{-5mm}
\section*{Data Availability}

The training dataset of DeepExtractor reconstructions used in this work, along with pretrained GlitchGAN generator weights, is hosted on Hugging Face and can be downloaded directly via the \texttt{glitchgan} Python package (\url{https://github.com/tomdooney95/glitchgan}).

\vspace{-5mm}

\appendix

\section{\label{sec:cDVGAN_Architecture}GlitchGAN Architecture}

Table \ref{tab:model_architecture_2} shows the cDVGAN architecture featured in GlitchGAN.

\begin{table}[H]
\centering
\captionsetup{width=0.5\textwidth}
\caption[Architecture and hyperparameters of cDVGAN featured in GlitchGAN]{Architecture and hyperparameters of cDVGAN featured in GlitchGAN, for the 8192-sample glitch waveforms used here (10.2M parameters total). The model includes a base discriminator, derivative (DV) discriminator, and generator.}
\label{tab:model_architecture_2}
\resizebox{0.43\textwidth}{!}{
\begin{tabular}{llllll}
\toprule
{} & {} & \textbf{Discriminator} & {} & {} & {(3.7M parameters)} \\ \hline
\textbf{Operation} & \textbf{Output shape} & \textbf{Kernel size} & \textbf{Stride} & \textbf{Dropout} & \textbf{Activation} \\ \hline
Input & (8192) & -- & -- & 0 & -- \\
Reshape & (64,128) & -- & -- & 0 & -- \\
Convolutional & (64,128) & 14 & 1 & 0.5 & Leaky ReLU \\
Convolutional & (32,128) & 14 & 2 & 0.5 & Leaky ReLU \\
Convolutional & (16,256) & 14 & 2 & 0.5 & Leaky ReLU \\
Convolutional & (8,256) & 14 & 2 & 0.5 & Leaky ReLU \\
Convolutional & (4,512) & 14 & 2 & 0.5 & Leaky ReLU \\
Global Avg. Pooling & (512) & -- & -- & 0 & -- \\
Dense & (128) & -- & -- & 0 & Linear \\
Dense & (1) & -- & -- & 0 & Linear \\
Class input & (7) & -- & -- & -- & -- \\
Class embedding & (128) & -- & -- & 0 & -- \\
Scalar product & (1) & -- & -- & -- & -- \\
Dense + Scalar product & (1) & -- & -- & -- & -- \\ \bottomrule

{} & {} & \textbf{DV Discriminator} & {} & {} & {(2.7M parameters)} \\ \hline
\textbf{Operation} & \textbf{Output shape} & \textbf{Kernel size} & \textbf{Stride} & \textbf{Dropout} & \textbf{Activation} \\ \hline
Input & (8191) & -- & -- & 0 & -- \\
Dense & (256) & -- & -- & 0 & Leaky ReLU \\
Reshape & (8,32) & -- & -- & 0 & -- \\
Convolutional & (8,64) & 5 & 1 & 0.5 & Leaky ReLU \\
Convolutional & (4,128) & 5 & 2 & 0.5 & Leaky ReLU \\
Convolutional & (2,256) & 5 & 2 & 0.5 & Leaky ReLU \\
Convolutional & (1,256) & 5 & 2 & 0.5 & Leaky ReLU \\
Global Avg. Pooling & (256) & -- & -- & 0 & -- \\
Dense & (128) & -- & -- & 0 & Linear \\
Dense & (1) & -- & -- & 0 & Linear \\
Class input & (7) & -- & -- & -- & -- \\
Class embedding & (128) & -- & -- & 0 & -- \\
Scalar product & (1) & -- & -- & -- & -- \\
Dense + Scalar product & (1) & -- & -- & -- & -- \\ \bottomrule

{} & {} & \textbf{Generator} & {} & {} & {(3.8M parameters)} \\ \hline
\textbf{Operation} & \textbf{Output shape} & \textbf{Kernel size} & \textbf{Stride} & \textbf{BN} & \textbf{Activation} \\ \hline
Latent input & (100) & -- & -- & \xmark & -- \\
Class input & (7) & -- & -- & \xmark & -- \\
Class Dense & (32) & -- & -- & \xmark & Linear \\
Concatenate & (132) & -- & -- & \xmark & -- \\
Dense & (4096) & -- & -- & \xmark & ReLU \\
Reshape & (256,16) & -- & -- & \xmark & -- \\
Upsample ($\times$2) + Conv. & (512,512) & 18 & 1 & \cmark & ReLU \\
Upsample ($\times$2) + Conv. & (1024,256) & 18 & 1 & \cmark & ReLU \\
Upsample ($\times$2) + Conv. & (2048,128) & 18 & 1 & \cmark & ReLU \\
Upsample ($\times$2) + Conv. & (4096,64) & 18 & 1 & \cmark & ReLU \\
Upsample ($\times$2) + Conv. & (8192,1) & 18 & 1 & \xmark & Linear \\
Flatten & (8192) & -- & -- & \xmark & -- \\ \hline

Optimizer & RMSprop ($\alpha = 0.0001$) & {} & {} & {} & {} \\
Batch size & 64 & {} & {} & {} & {} \\
Epochs & 210 & {} & {} & {} & {} \\
Loss & Wasserstein & {} & {} & {} & {} \\ \bottomrule
\end{tabular}
}
\end{table}


\section{\label{sec:cDVGAN_Loss}cDVGAN Loss}
\begin{figure}[ht]
\centering
\includegraphics[width=\linewidth]{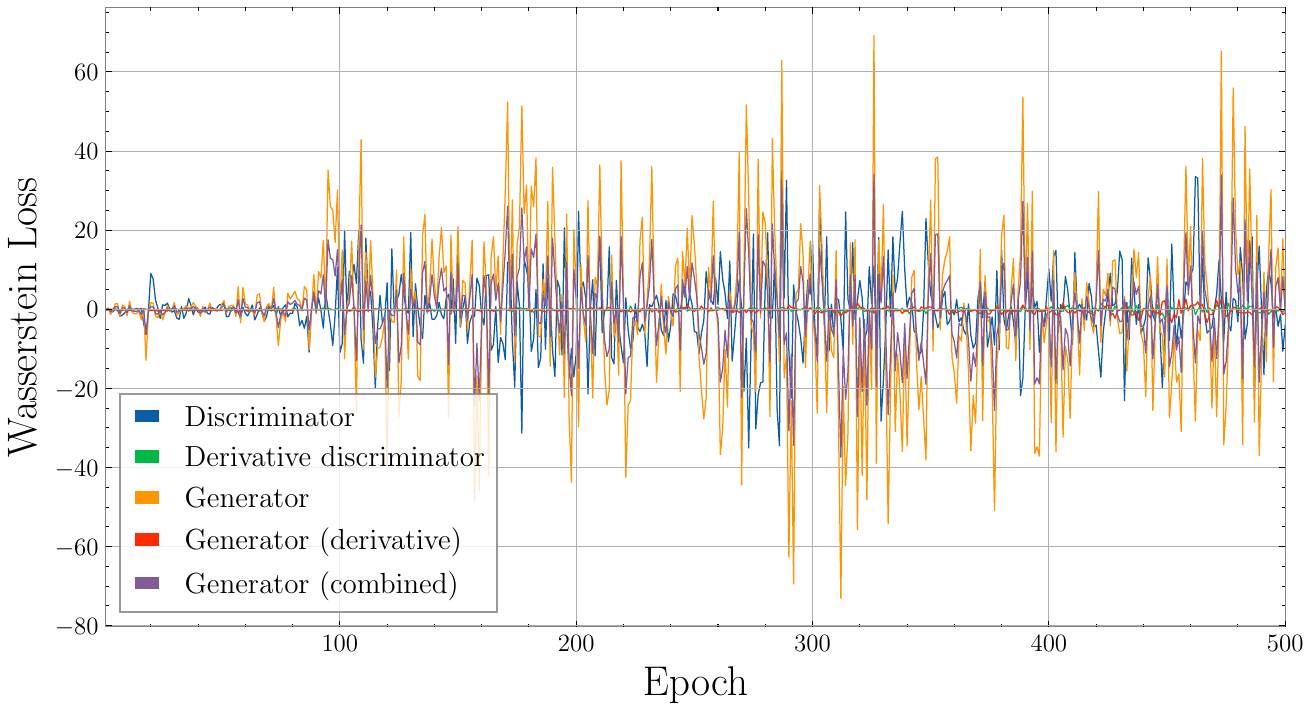}
\caption{GlitchGAN training loss over 500 epochs.}
\label{fig:loss_plot}
\end{figure}

Figure \ref{fig:loss_plot} shows the training loss for each individual model component.

\section{\label{sec:tsne_umap_params}Hyperparameters for t-SNE and UMAP}

Table \ref{tab:hyperparameters} lists the hyperparameters used for the t-SNE and UMAP analyses.

\begin{table}[h]
\centering
\caption[Hyperparameters for t-SNE and UMAP]{Hyperparameters for t-SNE and UMAP. For UMAP, \textbf{bold} values give the primary configuration used for the Figures~\protect\ref{fig:umap_two_views} and~\protect\ref{fig:umap_per_class_rotated}; the additional plain values were swept over to confirm the real/fake patterns are consistent and not an artifact of one particular embedding (9 total (\texttt{Random\_state}, \texttt{(N\_neighbors, Min\_dist)}) combinations: seeds $\{0,1,2\}$ $\times$ the three neighbor/distance pairs listed).}
\label{tab:hyperparameters}
\resizebox{0.43\textwidth}{!}{
\begin{tabular}{lll}
\hline
\textbf{Algorithm} & \textbf{Hyperparameter} & \textbf{Value} \\ \hline
\rule{0pt}{2.5ex}\textbf{t-SNE} & \texttt{N\_components} & 3 \\
 & \texttt{Perplexity} & 40 \\
 & \texttt{Metric} & \texttt{euclidean} \\ \hline
\rule{0pt}{2.5ex}\textbf{UMAP} & \texttt{(N\_neighbors, Min\_dist)} & \textbf{(15, 0.6)}, (30, 0.3), (50, 0.1) \\
 & \texttt{Random\_state} & \textbf{42}, 0, 1, 2 \\
 & \texttt{N\_components} & 3 \\
 & \texttt{Metric} & \texttt{correlation} \\ \hline
\end{tabular}}
\end{table}

\section{\label{sec:downstream_classifier_architectures}Downstream Classifier Architectures}

Tables~\ref{tab:real_fake_clf_arch} and~\ref{tab:multiclass_clf_arch} show the architectures of the binary real-vs-generated classifier (Section~\ref{sec:downstream_classification_methods}) and the seven-class multi-class classifier used for the data-augmentation experiment (Section~\ref{sec:multiclass_methods}), respectively. 
Figure~\ref{fig:snr_distributions} shows the empirical per-class SNR distributions used to inject both real and generated glitches into realistic detector noise for both experiments.

\begin{table}[h]
\centering
\captionsetup{width=0.5\textwidth}
\caption{Architecture of the 1D CNN classifier used to discriminate between real and GlitchGAN-generated glitches in the downstream classification test. The network takes an 8192-sample time series as input and outputs a single scalar in $[0,1]$, trained with a binary cross-entropy loss and Adam optimizerto predict real (1) vs.\ generated (0). Total trainable parameters: 172{,}897.}
\label{tab:real_fake_clf_arch}
\resizebox{0.43\textwidth}{!}{
\begin{tabular}{llllll}
\toprule
Operation & Output shape & Kernel size & Stride & Dropout or BN & Activation \\
\midrule
Input                    & (8192, 1)  & --  & -- & --  & --      \\
Conv1D                   & (2048, 32) & 16  & 4  & --  & ReLU    \\
Conv1D                   & (512, 64)  & 16  & 4  & --  & ReLU    \\
Conv1D                   & (128, 128) & 16  & 4  & --  & ReLU    \\
GlobalAveragePooling1D   & (128,)     & --  & -- & --  & --      \\
Dense                    & (64,)      & --  & -- & --  & ReLU    \\
Dropout                  & (64,)      & --  & -- & 0.3 & --      \\
Dense                    & (1,)       & --  & -- & --  & Sigmoid \\
\bottomrule
\end{tabular}}
\end{table}

\begin{table}[h]
\centering
\captionsetup{width=0.5\textwidth}
\caption[Architecture of the seven-class CNN classifier for multi-class data augmentation]{Architecture of the seven-class CNN classifier used for the multi-class data-augmentation experiment (Section~\protect\ref{sec:multiclass_methods}). Identical convolutional backbone to the binary real-vs-generated classifier (Table~\protect\ref{tab:real_fake_clf_arch}), differing only in the final layer (seven-way softmax in place of a single sigmoid) and loss (sparse categorical cross-entropy). Total trainable parameters: 173{,}287.}
\label{tab:multiclass_clf_arch}
\resizebox{0.43\textwidth}{!}{
\begin{tabular}{llllll}
\toprule
Operation & Output shape & Kernel size & Stride & Dropout or BN & Activation \\
\midrule
Input                    & (8192, 1)  & --  & -- & --  & --      \\
Conv1D                   & (2048, 32) & 16  & 4  & --  & ReLU    \\
Conv1D                   & (512, 64)  & 16  & 4  & --  & ReLU    \\
Conv1D                   & (128, 128) & 16  & 4  & --  & ReLU    \\
GlobalAveragePooling1D   & (128,)     & --  & -- & --  & --      \\
Dense                    & (64,)      & --  & -- & --  & ReLU    \\
Dropout                  & (64,)      & --  & -- & 0.3 & --      \\
Dense                    & (7,)       & --  & -- & --  & Softmax \\
\bottomrule
\end{tabular}}
\end{table}

\vspace{-2mm}

\section{\label{sec:snr_distributions}Empirical SNR distributions for noise injection}

\begin{figure}[h]
\centering
\includegraphics[width=\linewidth]{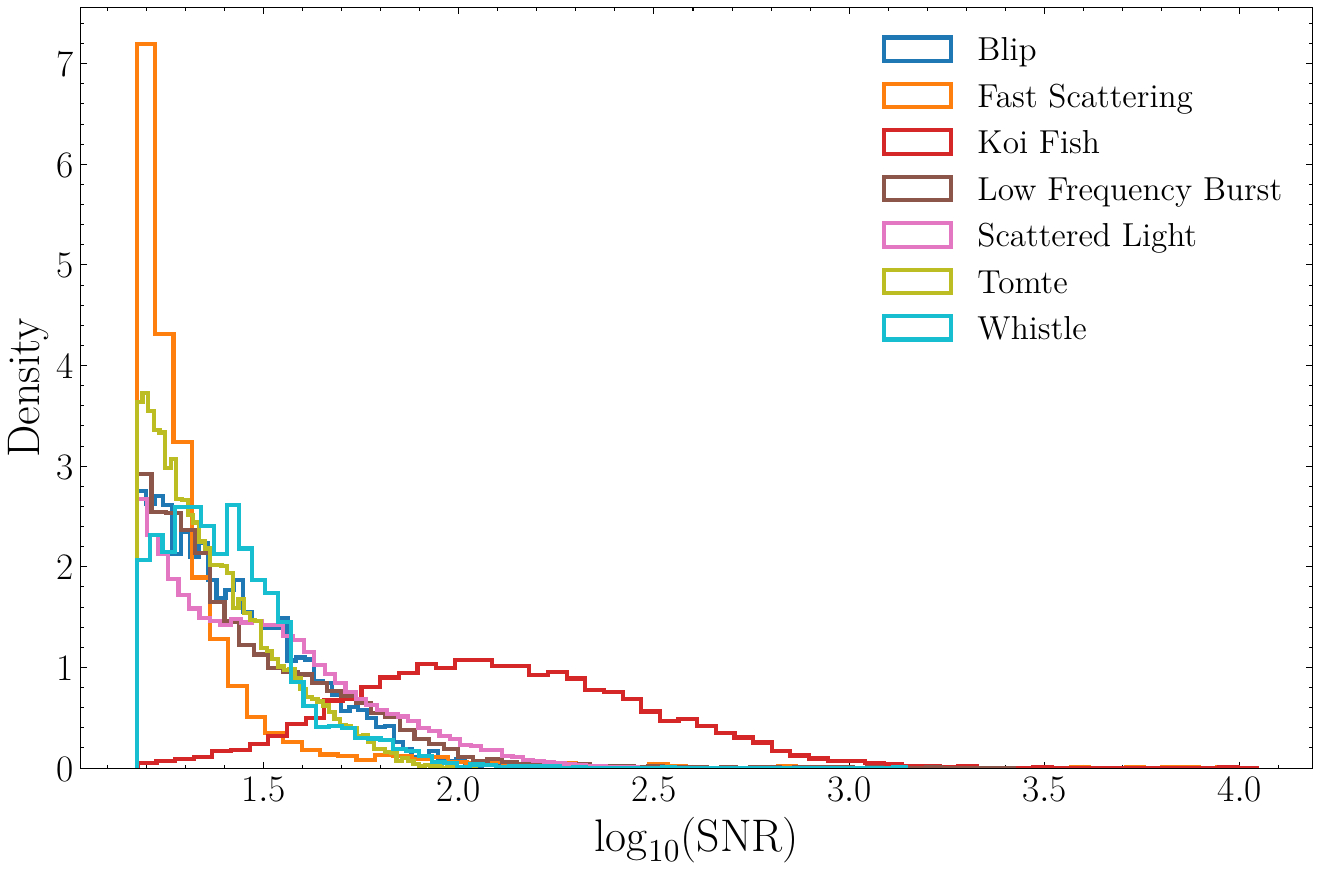}
\caption[Empirical per-class SNR distributions used for noise injection]{Empirical per-class SNR distributions used for realistic noise injection, constructed from the GravitySpy O3a/O3b high-confidence catalogs under the same selection criteria as the training dataset (Section~\protect\ref{sec:data}). Distributions are heavily right-skewed; note the log-scale horizontal axis. Koi Fish is a visibly distinct, systematically louder population than the other six classes.}
\label{fig:snr_distributions}
\end{figure}

Figure~\ref{fig:snr_distributions} shows the empirical per-class SNR distributions used to inject both real and generated glitches into realistic detector noise for the downstream classification experiments (Sections~\ref{sec:downstream_classification_methods} and~\ref{sec:multiclass_methods}).

\vspace{-2mm}
\section{DiT evaluation results}
\label{app:dit_evaluation}

\begin{figure*}[ht]
\centering
\includegraphics[width=0.9\linewidth]{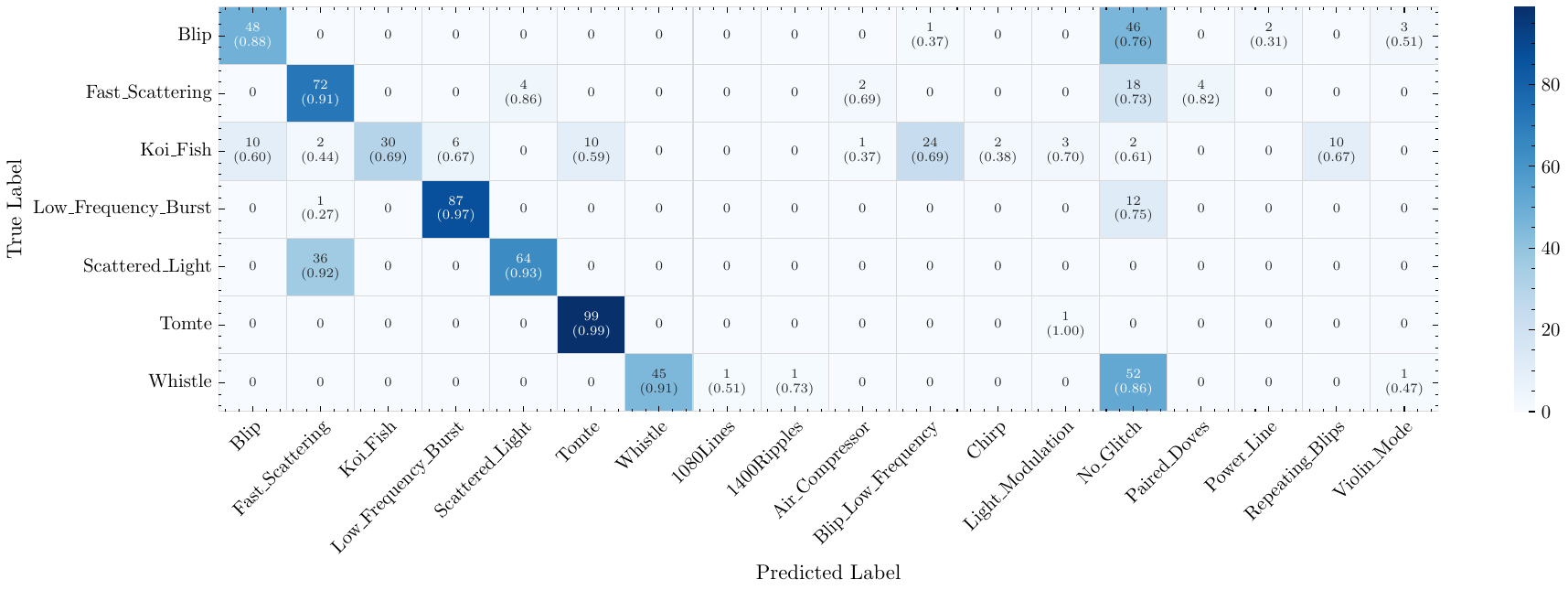}
\caption{Gravity Spy confusion matrix for samples generated by the DiT model (100 per class). Each cell shows the number of predictions and mean classifier confidence in brackets. Despite the poor UMAP overlap, Gravity Spy achieves high confidence classification for most classes, illustrating the limitation of magnitude-only validation.}
\label{fig:dit_confusion}
\end{figure*}

\begin{figure*}[ht]
\centering
\includegraphics[width=0.8\linewidth]{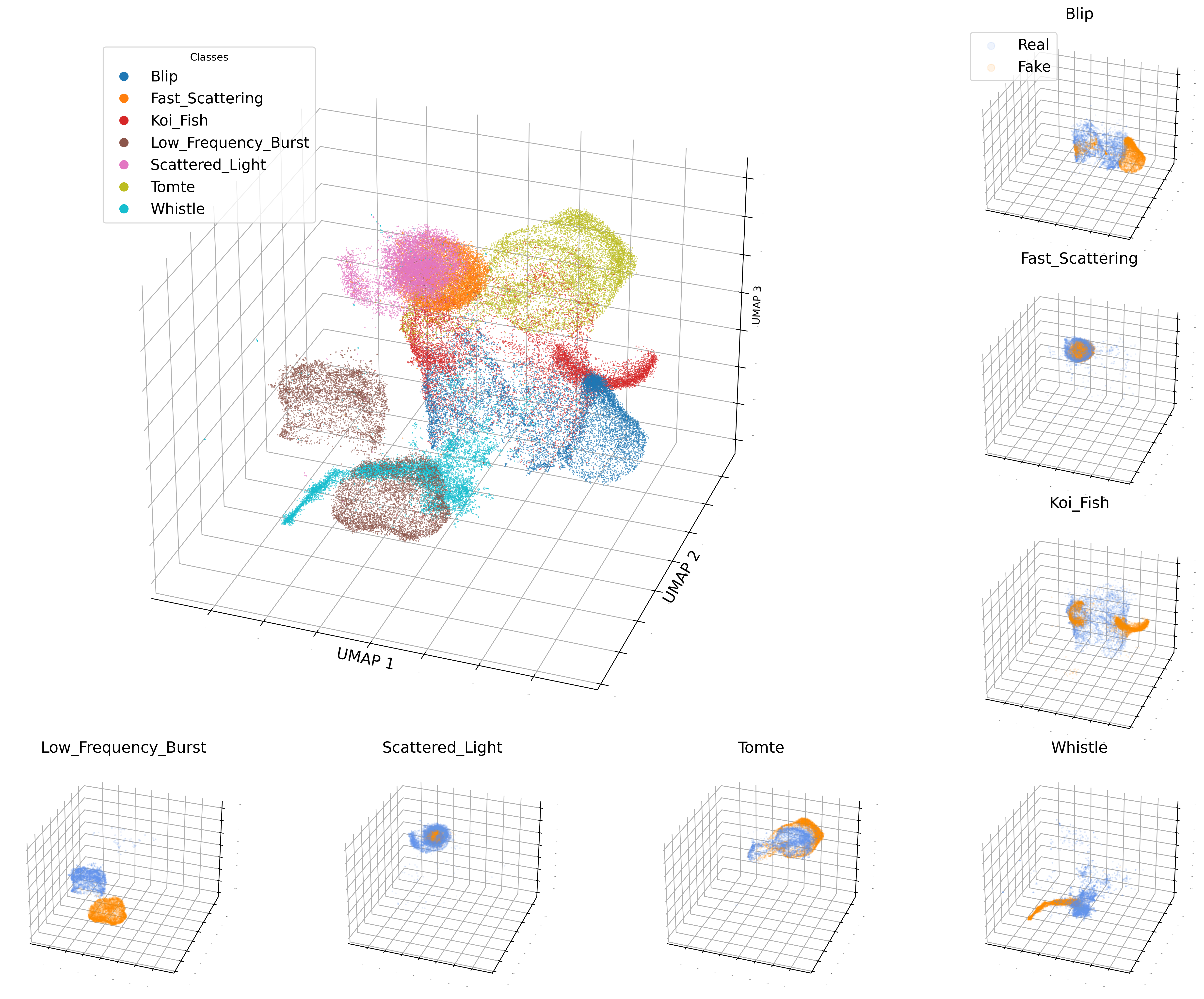}
\caption[Real vs.\ DiT-generated glitches in UMAP embedding space]{Three-dimensional UMAP projection of real (orange) and DiT-generated (blue) glitches across all seven classes. The large panel shows the overall structure, while smaller panels display each class individually. Overlapping clusters indicate morphological similarity between real and synthetic samples; separable clusters for most classes indicate that DiT does not reproduce the full morphological structure of real glitches, in contrast to GlitchGAN (Figure~\protect\ref{fig:umap_two_views}).}
\label{fig:dit_umap}
\end{figure*}

Figures~\ref{fig:dit_confusion} and~\ref{fig:dit_umap} show the full Gravity Spy and UMAP evaluation results for the DiT model.

\bibliography{references}

@article{Mirza_2014,
  author  = {Mirza, M. and Osindero, S.},
  title   = {Conditional Generative Adversarial Nets},
  journal = {arXiv preprint},
  year    = {2014},
  eprint  = {1411.1784},
  archivePrefix = {arXiv}
}

@article{McInnes_2018,
  author  = {McInnes, L. and Healy, J. and Melville, J.},
  title   = {{UMAP: Uniform Manifold Approximation and Projection for Dimension Reduction}},
  journal = {arXiv preprint},
  year    = {2018},
  eprint  = {1802.03426},
  archivePrefix = {arXiv}
}

@article{DeepExtractor_paper,
  author  = {Dooney, T. and others},
  title   = {Time-domain reconstruction of signals and glitches in gravitational wave data with deep learning},
  journal = {Physical Review D},
  year    = {2025},
  doi     = {10.1103/s91m-c2jw}
}

@misc{Areeda2017,
  author = {Areeda, Joseph},
  title = {Are Blips and Koi Fish Related?},
  howpublished = {\url{https://blog.gravityspy.org/2017/10/19/are-blips-and-koi-fish-related/}},
  year = {2017},
  month = {10},
  day = {19},
  note = {Accessed on 11-06-2025}
}

@article{LIGO_O3_detector_characterization,
  author  = {{The LIGO Scientific Collaboration} and others},
  title   = {Characterization of the LIGO detectors during their third observing run},
  journal = {Classical and Quantum Gravity},
  volume  = {38},
  pages   = {135003},
  year    = {2021},
  doi     = {10.1088/1361-6382/abf4c0}
}

@article{LIGO_paper,
       author = {{LIGO Scientific Collaboration and others}},
        title = "{Advanced LIGO}",
      journal = {Class. Quant. Grav.},
         year = 2015,
        month = apr,
       volume = {32},
       number = {7},
          eid = {074001},
        pages = {074001},
          doi = {10.1088/0264-9381/32/7/074001},
archivePrefix = {arXiv},
       eprint = {1411.4547},
 primaryClass = {gr-qc},
       adsurl = {https://ui.adsabs.harvard.edu/abs/2015CQGra..32g4001L}
}

@article{VIRGOpaper,
    author = "F. Acernese and others",
    collaboration = "VIRGO",
    title = "{Advanced Virgo: a second-generation interferometric gravitational wave detector}",
    eprint = "1408.3978",
    archivePrefix = "arXiv",
    primaryClass = "gr-qc",
    doi = "10.1088/0264-9381/32/2/024001",
    journal = "Class. Quant. Grav.",
    volume = "32",
    number = "2",
    pages = "024001",
    year = "2015"
}

@article{LIGO_second_third_detchar,
   title={LIGO detector characterization in the second and third observing runs},
   volume={38},
   ISSN={1361-6382},
   url={http://dx.doi.org/10.1088/1361-6382/abfd85},
   DOI={10.1088/1361-6382/abfd85},
   number={13},
   journal={Classical and Quantum Gravity},
   publisher={IOP Publishing},
   author={Davis, D and others},
   year={2021},
   month=jun, pages={135014} }

@article{Virgo_detchar_O3,
    author = "Acernese, F. and others",
    collaboration = "Virgo",
    title = "{Virgo detector characterization and data quality: results from the O3 run}",
    eprint = "2210.15633",
    archivePrefix = "arXiv",
    primaryClass = "gr-qc",
    doi = "10.1088/1361-6382/acd92d",
    journal = "Class. Quant. Grav.",
    volume = "40",
    number = "18",
    pages = "185006",
    year = "2023"
}

@article{Kagra_detchara,
    author = {Akutsu, T and others},
    title = "{Overview of KAGRA: Calibration, detector characterization, physical environmental monitors, and the geophysics interferometer}",
    journal = {Progress of Theoretical and Experimental Physics},
    volume = {2021},
    number = {5},
    pages = {05A102},
    year = {2021},
    month = {02},
    issn = {2050-3911},
    doi = {10.1093/ptep/ptab018},
    url = {https://doi.org/10.1093/ptep/ptab018},
    eprint = {https://academic.oup.com/ptep/article-pdf/2021/5/05A102/38109702/ptab018.pdf},
}

@article{abbott2018effects,
    author = "B. P. Abbott and others",
    collaboration = "LIGO Scientific, Virgo",
    title = "{Effects of data quality vetoes on a search for compact binary coalescences in Advanced LIGO's first observing run}",
    eprint = "1710.02185",
    archivePrefix = "arXiv",
    primaryClass = "gr-qc",
    doi = "10.1088/1361-6382/aaaafa",
    journal = "Class. Quant. Grav.",
    volume = "35",
    number = "6",
    pages = "065010",
    year = "2018"
}

@article{pankow2018mitigation,
  title={Mitigation of the instrumental noise transient in gravitational-wave data surrounding GW170817},
  author={C. Pankow and others},
  journal={Phys. Rev. D},
  volume={98},
  number={8},
  pages={084016},
  year={2018},
  publisher={APS}
}

@article{blackburn2008lsc,
    author = "L. Blackburn and others",
    editor = "Hughes, S. and Katsavounidis, E.",
    title = "{The LSC Glitch Group: Monitoring Noise Transients during the fifth LIGO Science Run}",
    eprint = "0804.0800",
    archivePrefix = "arXiv",
    primaryClass = "gr-qc",
    reportNumber = "LIGO-P080016-00, LIGO-P080016-01",
    doi = "10.1088/0264-9381/25/18/184004",
    journal = "Class. Quant. Grav.",
    volume = "25",
    pages = "184004",
    year = "2008"
}

@article{abbott2016characterization,
    author = "B. P. Abbott and others",
    collaboration = "LIGO Scientific, Virgo",
    title = "{Characterization of transient noise in Advanced LIGO relevant to gravitational wave signal GW150914}",
    eprint = "1602.03844",
    archivePrefix = "arXiv",
    primaryClass = "gr-qc",
    doi = "10.1088/0264-9381/33/13/134001",
    journal = "Class. Quant. Grav.",
    volume = "33",
    number = "13",
    pages = "134001",
    year = "2016"
}

@article{GWTC_3_catalog,
   title={GWTC-3: Compact Binary Coalescences Observed by LIGO and Virgo during the Second Part of the Third Observing Run},
   volume={13},
   ISSN={2160-3308},
   url={http://dx.doi.org/10.1103/PhysRevX.13.041039},
   DOI={10.1103/physrevx.13.041039},
   number={4},
   journal={Physical Review X},
   publisher={American Physical Society (APS)},
   author={Abbott, R. and others},
   year={2023},
   month=dec }

@article{BNS_1,
  title = {GW170817: Observation of Gravitational Waves from a Binary Neutron Star Inspiral},
  author = {Abbott, B. P. and others},
  collaboration = {LIGO Scientific Collaboration and Virgo Collaboration},
  journal = {Phys. Rev. Lett.},
  volume = {119},
  issue = {16},
  pages = {161101},
  numpages = {18},
  year = {2017},
  month = {Oct},
  publisher = {American Physical Society},
  doi = {10.1103/PhysRevLett.119.161101},
  url = {https://link.aps.org/doi/10.1103/PhysRevLett.119.161101}
}

@article{BNS_2,
doi = {10.3847/2041-8213/ab75f5},
url = {https://dx.doi.org/10.3847/2041-8213/ab75f5},
year = {2020},
month = {mar},
publisher = {The American Astronomical Society},
volume = {892},
number = {1},
pages = {L3},
author = {B. P. Abbott and others},
title = {GW190425: Observation of a Compact Binary Coalescence with Total Mass $\sim 3.4 \, M_{\odot}$},
journal = {The Astrophysical Journal Letters}
}

@article{BNS_BBH_2,
doi = {10.3847/2041-8213/ac082e},
url = {https://dx.doi.org/10.3847/2041-8213/ac082e},
year = {2021},
month = {jun},
publisher = {The American Astronomical Society},
volume = {915},
number = {1},
pages = {L5},
author = {R. Abbott and others},
title = {Observation of Gravitational Waves from Two Neutron Star–Black Hole Coalescences},
journal = {The Astrophysical Journal Letters}
}

@article{cdvgan,
  title = {One flexible model for multiclass gravitational wave signal and glitch generation},
  author = {Dooney, Tom and Curier, R. Lyana and Tan, Daniel Stanley and Lopez, Melissa and Van Den Broeck, Chris and Bromuri, Stefano},
  journal = {Phys. Rev. D},
  volume = {110},
  issue = {2},
  pages = {022004},
  numpages = {19},
  year = {2024},
  month = {7},
  publisher = {American Physical Society},
  doi = {10.1103/PhysRevD.110.022004},
  url = {https://link.aps.org/doi/10.1103/PhysRevD.110.022004}
}

@article{Jade_Powell_paper,
doi = {10.1088/1361-6382/acb038},
url = {https://doi.org/10.1088/1361-6382/acb038},
year = {2023},
month = {1},
publisher = {IOP Publishing},
volume = {40},
number = {3},
pages = {035006},
author = {Powell, Jade and Sun, Ling and Gereb, Katinka and Lasky, Paul D and Dollmann, Markus},
title = {Generating transient noise artefacts in gravitational-wave detector data with generative adversarial networks},
journal = {Classical and Quantum Gravity}
}

@article{PE_glitch_1,
  title = {Curious case of GW200129: Interplay between spin-precession inference and data-quality issues},
  author = {Payne, Ethan and Hourihane, Sophie and Golomb, Jacob and Udall, Rhiannon and Davis, Derek and Chatziioannou, Katerina},
  journal = {Phys. Rev. D},
  volume = {106},
  issue = {10},
  pages = {104017},
  numpages = {18},
  year = {2022},
  month = {11},
  publisher = {American Physical Society},
  doi = {10.1103/PhysRevD.106.104017},
  url = {https://link.aps.org/doi/10.1103/PhysRevD.106.104017}
}

@misc{PE_glitch_2,
      title={Possible Causes of False General Relativity Violations in Gravitational Wave Observations}, 
      author={Anuradha Gupta and others},
      year={2024},
      eprint={2405.02197},
      archivePrefix={arXiv},
      primaryClass={gr-qc},
      url={https://arxiv.org/abs/2405.02197}, 
}

@INPROCEEDINGS{dooney2022dvgan,
  author={Dooney, Tom and Bromuri, Stefano and Curier, Lyana},
  booktitle={2022 IEEE International Conference on Big Data (Big Data)}, 
  title={DVGAN: Stabilize {W}asserstein {GAN} training for time-domain Gravitational Wave physics}, 
  year={2022},
  volume={},
  number={},
  pages={5468-5477},
  address={Osaka, Japan},
  doi={10.1109/BigData55660.2022.10021080}}

@article{BayesWave,
	doi = {10.1088/0264-9381/32/13/135012},
	url = {https://doi.org/10.1088/0264-9381/32/13/135012},
	year = 2015,
	month = {jun},
	publisher = {{IOP} Publishing},
	volume = {32},
	number = {13},
	pages = {135012},
	author = {Neil J Cornish and Tyson B Littenberg},
	title = {Bayeswave: Bayesian inference for gravitational wave bursts and instrument glitches},
	journal = {Class. Quant. Grav.},
}

@InProceedings{wGAN_paper,
  title = 	 {{W}asserstein Generative Adversarial Networks},
  author =       {Martin Arjovsky and Soumith Chintala and L{\'e}on Bottou},
  booktitle = 	 {Proceedings of the 34th International Conference on Machine Learning},
  pages = 	 {214--223},
  year = 	 {2017},
  editor = 	 {Precup, Doina and Teh, Yee Whye},
  volume = 	 {70},
  series = 	 {Proceedings of Machine Learning Research},
  month = 	 {06--11 Aug},
  publisher =    {PMLR},
  url = 	 {https://proceedings.mlr.press/v70/arjovsky17a.html},
}

@article{McGinn_2021,
	doi = {10.1088/1361-6382/ac09cc},
  
	url = {https://doi.org/10.1088\%2F1361-6382\%2Fac09cc},
  
	year = 2021,
	month = {jun},
  
	publisher = {{IOP} Publishing},
  
	volume = {38},
  
	number = {15},
  
	pages = {155005},
  
	author = {J McGinn and others},
  
	title = {Generalised gravitational wave burst generation with generative adversarial networks},
  
	journal = {Class. Quant. Grav.}
}

@misc{GENGLI,
  doi = {10.48550/ARXIV.2203.06494},
  url = {https://arxiv.org/abs/2203.06494},
  author = {M. Lopez and others},
  title = {Simulating Transient Noise Bursts in LIGO with Generative Adversarial Networks},
  publisher = {arXiv},
  year = {2022},
  copyright = {arXiv.org perpetual, non-exclusive license}
}

@article{Zevin_2017,
	doi = {10.1088/1361-6382/aa5cea},
  
	url = {https://doi.org/10.1088\%2F1361-6382\%2Faa5cea},
  
	year = 2017,
	month = {feb},
  
	publisher = {{IOP} Publishing},
  
	volume = {34},
  
	number = {6},
  
	pages = {064003},
  
	author = {M Zevin and others},
  
	title = {Gravity Spy: integrating advanced {LIGO} detector characterization, machine learning, and citizen science},
  
	journal = {Class. Quant. Grav.}
}

@misc{Goodfellow:2014upx,
    author = "Goodfellow, I. J. and others",
    title = "{Generative Adversarial Networks}",
    eprint = "1406.2661",
    archivePrefix = "arXiv",
    primaryClass = "stat.ML",
    month = "6",
    year = "2014"
}

@misc{wGAN_GP_paper,
  doi = {10.48550/ARXIV.1704.00028},
  url = {https://arxiv.org/abs/1704.00028},
  author = {Gulrajani, Ishaan and Ahmed, Faruk and Arjovsky, Martin and Dumoulin, Vincent and Courville, Aaron},
  title = {Improved Training of Wasserstein GANs},
  publisher = {arXiv},
  year = {2017},
  copyright = {arXiv.org perpetual, non-exclusive license}
}

@article{Liao:2021vec,
    author = "C-H. Liao and F-L. Lin",
    title = "{Deep generative models of gravitational waveforms via conditional autoencoder}",
    eprint = "2101.06685",
    archivePrefix = "arXiv",
    primaryClass = "astro-ph.IM",
    doi = "10.1103/PhysRevD.103.124051",
    journal = "Phys. Rev. D",
    volume = "103",
    number = "12",
    pages = "124051",
    year = "2021"
}

@article{PhysRevD.110.104055,
  title = {Generative adversarial network for stellar core-collapse gravitational waves},
  author = {Eccleston, Tarin and Edwards, Matthew C.},
  journal = {Phys. Rev. D},
  volume = {110},
  issue = {10},
  pages = {104055},
  numpages = {11},
  year = {2024},
  month = {11},
  publisher = {American Physical Society},
  doi = {10.1103/PhysRevD.110.104055}
}

@misc{GAN_spec_best,
  doi = {10.48550/ARXIV.2207.04001},
  url = {https://arxiv.org/abs/2207.04001},
  author = {Yan, Jianqi and Leung, Alex P. and Hui, David C. Y.},
  title = {On Improving the Performance of Glitch Classification for Gravitational Wave Detection by using Generative Adversarial Networks},
  publisher = {arXiv},
  year = {2022},
  copyright = {arXiv.org perpetual, non-exclusive license}
}

@misc{GENGLI_2,
  doi = {10.48550/ARXIV.2205.09204},
  url = {https://arxiv.org/abs/2205.09204},
  author = {Lopez, Melissa and Boudart, Vincent and Schmidt, Stefano and Caudill, Sarah},
  title = {Simulating Transient Noise Bursts in LIGO with gengli},
  publisher = {arXiv},
  year = {2022},
  copyright = {arXiv.org perpetual, non-exclusive license}
}

@article{Abadie_2010_injection_1,
   title={Search for gravitational waves from compact binary coalescence in LIGO and Virgo data from S5 and VSR1},
   volume={82},
   ISSN={1550-2368},
   url={http://dx.doi.org/10.1103/PhysRevD.82.102001},
   DOI={10.1103/physrevd.82.102001},
   number={10},
   journal={Physical Review D},
   publisher={American Physical Society (APS)},
   author={Abadie, J. and others},
   year={2010},
   month=nov }

@article{Abadie_2010_injection_2,
   title={All-sky search for gravitational-wave bursts in the first joint LIGO-GEO-Virgo run},
   volume={81},
   ISSN={1550-2368},
   url={http://dx.doi.org/10.1103/PhysRevD.81.102001},
   DOI={10.1103/physrevd.81.102001},
   number={10},
   journal={Physical Review D},
   publisher={American Physical Society (APS)},
   author={Abadie, J. and others},
   year={2010},
   month=may }

@misc{diffusionOriginal,
  doi = {10.48550/ARXIV.1503.03585},
  url = {https://arxiv.org/abs/1503.03585},
  author = {Sohl-Dickstein, Jascha and Weiss, Eric A. and Maheswaranathan, Niru and Ganguli, Surya},
  title = {Deep Unsupervised Learning using Nonequilibrium Thermodynamics},
  publisher = {arXiv},
  year = {2015},
  copyright = {arXiv.org perpetual, non-exclusive license}
}

@misc{GspyO4,
      title={Advancing Glitch Classification in Gravity Spy: Multi-view Fusion with Attention-based Machine Learning for Advanced LIGO's Fourth Observing Run}, 
      author={Yunan Wu and Michael Zevin and Christopher P. L. Berry and Kevin Crowston and Carsten Østerlund and Zoheyr Doctor and Sharan Banagiri and Corey B. Jackson and Vicky Kalogera and Aggelos K. Katsaggelos},
      year={2024},
      eprint={2401.12913},
      archivePrefix={arXiv},
      primaryClass={gr-qc}
}

@article{ET_paper,
	doi = {10.1088/0264-9381/28/9/094013},
  
	url = {https://doi.org/10.1088%2F0264-9381%2F28%2F9%2F094013},
  
	year = 2011,
	month = {apr},
  
	publisher = {{IOP} Publishing},
  
	volume = {28},
  
	number = {9},
  
	pages = {094013},
  
author = {S Hild and M Abernathy and F Acernese and P Amaro-Seoane and N Andersson and K Arun and F Barone and B Barr and M Barsuglia and M Beker e},
  
	title = {Sensitivity studies for third-generation gravitational wave observatories},
  
	journal = {Class. Quant. Grav.}
}

@Article{Cosmic_expl,
AUTHOR = {Hall , Evan D.},
TITLE = {Cosmic Explorer: A Next-Generation Ground-Based Gravitational-Wave Observatory},
JOURNAL = {Galaxies},
VOLUME = {10},
YEAR = {2022},
NUMBER = {4},
ARTICLE-NUMBER = {90},
URL = {https://www.mdpi.com/2075-4434/10/4/90},
ISSN = {2075-4434},
DOI = {10.3390/galaxies10040090}
}

@article{Cabero:2019orq,
    author = "M. Cabero and others",
    title = "{Blip glitches in Advanced LIGO data}",
    eprint = "1901.05093",
    archivePrefix = "arXiv",
    primaryClass = "physics.ins-det",
    doi = "10.1088/1361-6382/ab2e14",
    journal = "Class. Quant. Grav.",
    volume = "36",
    number = "15",
    pages = "15",
    year = "2019"
}

@article{Soni:2021cjy,
    author = "S. Soni and others",
    title = "{Discovering features in gravitational-wave data through detector characterization, citizen science and machine learning}",
    eprint = "2103.12104",
    archivePrefix = "arXiv",
    primaryClass = "gr-qc",
    doi = "10.1088/1361-6382/ac1ccb",
    journal = "Class. Quant. Grav.",
    volume = "38",
    number = "19",
    pages = "195016",
    year = "2021"
}

@article{Nuttall:2015dqa,
    author = "Nuttall, L. and others",
    title = "{Improving the Data Quality of Advanced LIGO Based on Early Engineering Run Results}",
    eprint = "1508.07316",
    archivePrefix = "arXiv",
    primaryClass = "gr-qc",
    doi = "10.1088/0264-9381/32/24/245005",
    journal = "Class. Quant. Grav.",
    volume = "32",
    number = "24",
    pages = "245005",
    year = "2015"
}

@misc{Ho2020DDPM,
      title={Denoising Diffusion Probabilistic Models}, 
      author={Jonathan Ho and Ajay Jain and Pieter Abbeel},
      year={2020},
      eprint={2006.11239},
      archivePrefix={arXiv},
      primaryClass={cs.LG},
      url={https://arxiv.org/abs/2006.11239}, 
}

@misc{ho2022classifierfree,
      title={Classifier-Free Diffusion Guidance}, 
      author={Jonathan Ho and Tim Salimans},
      year={2022},
      eprint={2207.12598},
      archivePrefix={arXiv},
      primaryClass={cs.LG},
      url={https://arxiv.org/abs/2207.12598}, 
}

@article{Bondarescu:2023jcx,
    author = "Bondarescu, Ruxandra and Lundgren, Andrew and Macas, Ronaldas",
    title = "{Quasiphysical model for removing short glitches from LIGO and Virgo data}",
    eprint = "2309.06594",
    archivePrefix = "arXiv",
    primaryClass = "gr-qc",
    doi = "10.1103/PhysRevD.108.122004",
    journal = "Phys. Rev. D",
    volume = "108",
    number = "12",
    pages = "122004",
    year = "2023"
}

@article{Tolley:2023umc,
    author = "A. E. Tolley and others",
    title = "{ArchEnemy: removing scattered-light glitches from gravitational wave data}",
    eprint = "2301.10491",
    archivePrefix = "arXiv",
    primaryClass = "gr-qc",
    doi = "10.1088/1361-6382/ace22f",
    journal = "Class. Quant. Grav.",
    volume = "40",
    number = "16",
    pages = "165005",
    year = "2023"
}

@misc{TSNE,
      title={Theoretical Foundations of t-SNE for Visualizing High-Dimensional Clustered Data}, 
      author={T. Tony Cai and Rong Ma},
      year={2022},
      eprint={2105.07536},
      archivePrefix={arXiv},
      primaryClass={stat.ML},
      url={https://arxiv.org/abs/2105.07536}, 
}

@misc{Umap,
      title={UMAP: Uniform Manifold Approximation and Projection for Dimension Reduction}, 
      author={Leland McInnes and John Healy and James Melville},
      year={2020},
      eprint={1802.03426},
      archivePrefix={arXiv},
      primaryClass={stat.ML},
      url={https://arxiv.org/abs/1802.03426}, 
}

@misc{tensorflow2015-whitepaper,
title={ {TensorFlow}: Large-Scale Machine Learning on Heterogeneous Systems},
url={http://tensorflow.org/},
note={Software available from tensorflow.org},
author={
    Mart\'{\i}n~Abadi and
    others},
  year={2015},
}

@article{Iacovelli_2022,
	doi = {10.3847/1538-4357/ac9cd4},
	url = {https://doi.org/10.3847%2F1538-4357%2Fac9cd4},
	year = 2022,
	month = {12},
	publisher = {American Astronomical Society},
	volume = {941},
	number = {2},
	pages = {208},
	author = {Francesco Iacovelli and others},
	title = {Forecasting the Detection Capabilities of Third-generation Gravitational-wave Detectors Using {GWFAST}
},
	journal = {The Astrophysical Journal}
}

@inproceedings{Kalogera_BNS,
  title={The Next Generation Global Gravitational Wave Observatory: The Science Book},
  author={Vicky Kalogera and others},
  year={2021},
  url={https://api.semanticscholar.org/CorpusID:244117472}
}

@misc{davis2026rapiddataqualityinvestigations,
      title={Rapid data quality investigations of gravitational-wave events with the Data Quality Report Builder toolkit}, 
      author={Derek Davis and others},
      year={2026},
      eprint={2605.16183},
      archivePrefix={arXiv},
      primaryClass={astro-ph.IM},
      url={https://arxiv.org/abs/2605.16183}, 
}

@article{Glanzer_2023,
   title={Data quality up to the third observing run of advanced LIGO: Gravity Spy glitch classifications},
   volume={40},
   ISSN={1361-6382},
   url={http://dx.doi.org/10.1088/1361-6382/acb633},
   DOI={10.1088/1361-6382/acb633},
   number={6},
   journal={Classical and Quantum Gravity},
   publisher={IOP Publishing},
   author={Glanzer, J and Banagiri, S and Coughlin, S B and Soni, S and Zevin, M and Berry, C P L and Patane, O and Bahaadini, S and Rohani, N and Crowston, K and Kalogera, V and Østerlund, C and Trouille, L and Katsaggelos, A},
   year={2023},
   month=Feb, pages={065004} }

@article{Chatterjee_2025,
doi = {10.3847/2041-8213/ae1a5f},
url = {https://doi.org/10.3847/2041-8213/ae1a5f},
year = {2025},
month = {dec},
publisher = {The American Astronomical Society},
volume = {995},
number = {1},
pages = {L6},
author = {Chatterjee, Chayan and McGowan, Kaylah and Deshmukh, Suyash and Tyler-Howard, Nicholas and Jani, Karan},
title = {Machine Learning Confirms GW231123 is a “Lite” Intermediate Mass Black Hole Merger},
journal = {The Astrophysical Journal Letters}
}

@misc{ferreira2025analysisligoglitchesusing,
      title={An Analysis of LIGO Glitches Using t-SNE During the First Part of the Fourth LIGO-Virgo-KAGRA Observing Run}, 
      author={Tabata Aira Ferreira and Gabriela González and Osvaldo Salas},
      year={2025},
      eprint={2512.03440},
      archivePrefix={arXiv},
      primaryClass={astro-ph.IM},
      url={https://arxiv.org/abs/2512.03440}, 
}

@article{Ferreira_2025,
doi = {10.1088/1361-6382/add3b5},
url = {https://doi.org/10.1088/1361-6382/add3b5},
year = {2025},
month = {may},
publisher = {IOP Publishing},
volume = {42},
number = {10},
pages = {105010},
author = {Ferreira, Tabata Aira and González, Gabriela},
title = {Using t-SNE for characterizing glitches in LIGO detectors},
journal = {Classical and Quantum Gravity}
}

@misc{deshmukh2026soundnoiseleveraginginductive,
      title={The Sound of Noise: Leveraging the Inductive Bias of Pre-trained Audio Transformers for Glitch Identification in LIGO}, 
      author={Suyash Deshmukh and Chayan Chatterjee and Abigail Petulante and Tabata Aira Ferreira and Karan Jani},
      year={2026},
      eprint={2601.20034},
      archivePrefix={arXiv},
      primaryClass={astro-ph.IM},
      url={https://arxiv.org/abs/2601.20034}, 
}

@article{gwpy,
    title = "{GWpy: A Python package for gravitational-wave astrophysics}",
   author = {{Macleod}, D.~M. and {Areeda}, J.~S. and {Coughlin}, S.~B. and {Massinger}, T.~J. and {Urban}, A.~L.},
  journal = {SoftwareX},
   volume = 13,
    pages = 100657,
     year = 2021,
     issn = {2352-7110},
      doi = {10.1016/j.softx.2021.100657},
      url = {https://www.sciencedirect.com/science/article/pii/S2352711021000029},
}

@article{Wilson01061927,
author = {Edwin B. Wilson},
title = {Probable Inference, the Law of Succession, and Statistical Inference},
journal = {Journal of the American Statistical Association},
volume = {22},
number = {158},
pages = {209--212},
year = {1927},
publisher = {Taylor \& Francis},
doi = {10.1080/01621459.1927.10502953},


URL = { 
    
    
        https://www.tandfonline.com/doi/abs/10.1080/01621459.1927.10502953
    

},
eprint = { 
    
    
        https://www.tandfonline.com/doi/pdf/10.1080/01621459.1927.10502953
    

}

}

@inproceedings{wang2020cnngenerated,
  title     = {{CNN}-Generated Images Are Surprisingly Easy to Spot... for Now},
  author    = {Wang, Sheng-Yu and Wang, Oliver and Zhang, Richard and Owens, Andrew and Efros, Alexei A.},
  booktitle = {Proceedings of the IEEE/CVF Conference on Computer Vision and Pattern Recognition (CVPR)},
  pages     = {8695--8704},
  year      = {2020}
}

\end{document}